\documentclass[11pt]{article}
\pdfoutput=1

\usepackage{euscript}
\usepackage{amssymb}
\usepackage{amsfonts}
\usepackage{amsbsy}
\usepackage{epsfig}
\usepackage{amsthm}
\usepackage{amscd}
\usepackage{amstext}
\usepackage{verbatim}
\usepackage{amsmath}
\usepackage{cancel}
\usepackage{capt-of}
\usepackage{empheq}
\usepackage{subfigure}
\usepackage{xcolor}

\usepackage[pdftex]{hyperref}
\hypersetup{ 
colorlinks=true, 
linkcolor=black, 
citecolor=red, 
}

\def\ben{\begin{equation}}
\def\een{\end{equation}}
\def\half{{\textstyle{\frac{1}{2}}}}
\let\a=\alpha \let\b=\beta   
   
     \let\r=\rho

\let\w=\omega

\let\pa=\partial
\def\be{\begin{equation}}
\def\ee{\end{equation}}
\def\beq{\begin{equation}}
\def\eeq{\end{equation}}
\def\ba{\begin{array}}
\def\ea{\end{array}}

\def\dalemb#1#2{{\vbox{\hrule height .#2pt
       \hbox{\vrule width.#2pt height#1pt \kern#1pt
               \vrule width.#2pt}
       \hrule height.#2pt}}}

\newcommand{\bea}{\begin{eqnarray}}
\newcommand{\eea}{\end{eqnarray}}

\def\vep{{\varepsilon}}

\makeatletter
\newcommand*\bigcdot{\mathpalette\bigcdot@{.5}}
\newcommand*\bigcdot@[2]{\mathbin{\vcenter{\hbox{\scalebox{#2}{$\m@th#1\bullet$}}}}}
\makeatother

\renewcommand{\eqref}[1]{(\ref{eq:#1})}

\def\Lag{{\mathcal{L}}}

\def\ocal{{\mathcal{O}}}

\begin{document}
\pagenumbering{gobble}
\begin{flushright}
SU-ITP-16/04
\end{flushright}

\begin{center}

{ \Large {\bf
Hydrodynamic theory of \\ quantum fluctuating superconductivity
}}

\vspace{1cm}

Richard A. Davison,$^{1}$ Luca V. Delacr\'etaz,$^{2}$ Blaise Gout\'eraux$^{2,3}$ and Sean A. Hartnoll$^{2}$

\vspace{1cm}

{\small
$^{1}${\it Department of Physics, Harvard University, \\
Cambridge, Massachusetts, 02138, USA}}

\vspace{0.5cm}

{\small
$^{2}${\it Department of Physics, Stanford University, \\
Stanford, CA 94305-4060, USA }}

\vspace{0.5cm}

{\small
$^{3}${\it APC, Universit\'e Paris 7, CNRS, CEA, \\ Observatoire de Paris, Sorbonne Paris Cit\'e, F-75205, Paris Cedex 13, France}}

\vspace{1.6cm}

\end{center}

\begin{abstract}

A hydrodynamic theory of transport in quantum mechanically phase-disordered superconductors
is possible when supercurrent relaxation can be treated as a slow process. We obtain general results
for the frequency-dependent conductivity of such a regime. With time-reversal invariance, the conductivity is characterized by a Drude-like peak, with width given by the supercurrent relaxation rate. Using the memory matrix
formalism, we obtain a formula for this width (and hence also the dc resistivity) when the supercurrent is relaxed by
short range density-density interactions. This leads to a new -- effective field theoretic and fully quantum -- derivation of a classic result on flux flow resistance. With strong breaking of time-reversal invariance, the optical conductivity exhibits what we call a `hydrodynamic supercyclotron' resonance. We obtain the frequency and decay rate of this resonance
for the case of supercurrent relaxation due to an emergent Chern-Simons gauge field. The supercurrent decay rate in this `topologically ordered superfluid vortex liquid' is determined by the conductivities of the normal fluid component, rather than the vortex core.

\end{abstract}

\pagebreak
\pagenumbering{arabic}
\tableofcontents

\section{Introduction}

While superconductivity is often correctly captured by mean field physics, fluctuations can be important, especially with reduced dimensionality. The effects of thermal superconducting fluctuations are largely understood; they are well-described within Ginzburg-Landau and -- in two spatial dimensions -- Berezinskii-Kosterlitz-Thouless theory \cite{tinkham, bkt, larkin}. The physics of quantum superconducting fluctuations, in contrast, presents theoretical challenges. Yet,
a theory of quantum fluctuating superconductivity is likely necessary to address important questions such as the existence of zero temperature metallic phases in two dimensions (for an overview of the challenges, see \cite{pp}).

Quantum mechanical effects can naturally lead to fluctuations in the phase of the superconducting order parameter. Quantum phase fluctuations will be the topic of this paper. The phase and the charge density are canonically conjugate variables. Therefore, if e.g. Coulombic interactions act to suppress charge density fluctuations, phase fluctuations will necessarily be induced due to the uncertainty principle $\Delta \rho \Delta \phi \gtrsim \hbar$. These can destroy long range phase coherence (early papers to emphasize this fact were \cite{t1,t2}). In this paper we will develop a theoretically controlled framework that realizes this intuition. We do this by working in a regime in which the supercurrent relaxation rate can be treated as a small parameter. In any case, it is only in this regime that the metallic nature of the state can be unambiguously characterized as phase-disordered superconductivity.

\subsection{Experimental motivation}
\label{sec:exp}

Phase fluctuations are expected to be generically important in systems with a small superfluid density such as organic and cuprate superconductors \cite{em1, bozovic, rourke}. Relatively inefficient Coulomb screening can increase the importance of quantum effects \cite{emery}, as can proximity to a Mott transition \cite{nam1, nam2}. The most dramatic and established appearance of quantum phase fluctuations, however, is in disordered thin films which we now discuss in more detail.

Disordered thin films undergo `superfluid-insulator' transitions as a function of magnetic field
\cite{MI1,MI2,MI3,MI4, MI5} or film thickness/disorder \cite{thick1, thick1a}. Of great interest for our purposes, in both cases intermediate metallic phases often exist between the superconducting and insulating phases. See \cite{MI4, MI5, MI5a, MI6, MI7} and \cite{thick1, thick2, thick3, thick4}, respectively. These metallic phases have a residual zero temperature resistivity that can be orders of magnitude smaller than the normal state resistivity. In this regime, at least, it is plausible that the transport is controlled by a slow relaxation rate that is distinct from the single-particle relaxation rate. A natural possibility is that it corresponds to a slow supercurrent relaxation due to quantum phase fluctuations. This will be the scenario studied in this paper.

In a magnetic field, disorder in thin films destabilizes the vortex lattice and leads to mobile vortices at any nonzero temperature \cite{LO}. Therefore, thin films that are `superconducting' in fact only have vanishing resistivity at $T=0$ \cite{FFH}. This is widely observed in the thin film references quoted above and also, for instance, in LSCO in sufficiently large magnetic fields \cite{vlad}. These phases are good candidates for our approach also, as they are metallic at arbitrarily low temperatures where quantum phase fluctuations may be important.

Direct evidence for quantum phase fluctuations in the intermediate metallic phase comes from measurements of the ac conductivity in weakly disordered two dimensional InO$_x$ films \cite{ac}. These observations were the immediate motivation for our work. Previous measurements, such as the magnetic field and temperature dependence of the Nernst effect \cite{behnia1, behnia2}, had established the importance of phase fluctuations in InO$_x$. The ac measurements, however, access the $T \to 0$ regime and furthermore directly reveal a long timescale. This timescale will be the essential building block of our theory.

In \cite{ac} the complex conductivity $\sigma(\omega,T)$ of a weakly disordered InO$_x$ film was measured as the system was driven from superconducting to (weakly) insulating behavior by varying a magnetic field. The data shows that while the low temperature, zero-frequency superfluid stiffness vanishes for magnetic fields above $B_\text{sm} \approx 3 \; \text{Tesla}$, weakly insulating behavior does not onset until the magnetic fields are larger than $B_\text{cross} \approx 7.5 \; \text{Tesla}$. In the intermediate magnetic field range, the films are metallic and, close to the superconducting phase, the ac conductivity is characterized by a sharp `Drude-like' zero-frequency Lorentzian peak. In the $T \to 0$ limit, the width of this peak tends to zero precisely as $B$ is lowered to $B_\text{sm}$. These measurements therefore directly obtain a long current relaxation timescale that continuously diverges at the onset of superconductivity. The natural (and possibly unique) interpretation of this timescale is that it is the lifetime of a supercurrent that decays due to phase incoherence.

The zero temperature Drude-like peak just described is not a conventional Drude peak. The width of the peak is directly connected to superfluid dynamics rather than disorder. In this paper we develop a theory of Drude-like peaks caused by quantum phase fluctuating superconductivity. In particular, we obtain formulae for the width of the peak and, consequently, for the finite dc conductivity. If the width remains finite at $T=0$, the theory describes a zero temperature metallic phase due to phase-fluctuating superconductivity.

A further candidate for a Drude-like peak due to phase fluctuating superconductivity is that observed in an organic molecular metal close to a Mott transition. Specifically, in $\kappa$-(BEDT-TTF)$_2$Cu[N(CN)$_2$]Cl$_{1-x}$Br$_x$, with $x=0.73$, a Drude-like peak is seen to emerge at $T \lesssim 50$K \cite{drudelike}. At precisely these temperatures a substantial, magnetic field dependent Nernst effect is also observed \cite{nam2}. While this peak has been interpreted in terms of `coherent quasiparticles', the evidence for superconducting phase fluctuations over the same temperature range may warrant a new look. Phase fluctuations in this family of organic superconductors are known to lead to quantum vortex liquids when placed in a magnetic field, even away from the Mott transition (i.e. in $\kappa$-(BEDT-TTF)$_2$Cu(NCS)$_2$ \cite{organ}).

\subsection{Summary of approach and results}

The starting point will be superfluid hydrodynamics. The hydrodynamic description is based on symmetries alone and therefore describes any superfluid or superconducting state. It is valid with or without the existence of long-lived quasiparticles. We will work in the `incoherent' limit \cite{Hartnoll:2014lpa}, which effectively means -- in the present context -- that we assume that supercurrent relaxation is parametrically slower than momentum relaxation. In this way our hydrodynamic theory contains superfluid velocity but not normal velocity as a variable. This limit seems to be relevant to the data in \cite{ac}, in which the width of the Drude-like peak is solely determined by supercurrent relaxation (as it is much narrower than the normal state Drude peak).

We proceed to partially `break' the hydrodynamic description by allowing for weak relaxation of the superfluid velocity (necessarily due to vortices). This is a situation that is tailor-made for `memory matrix' techniques \cite{forster}, that are built around long lived quantities. Using the memory matrix, we obtain an expression for the supercurrent relaxation rate starting from certain charge density interactions in the low energy effective Hamiltonian of the system. This is the step in which microscopic input is required and at which the quantum uncertainty relation $\Delta \rho \Delta \phi \gtrsim \hbar$ plays a role. Here $\rho$ is the charge density. The power of a hydrodynamic approach is that we do not need to know many details about the microscopic Hamiltonian. Supercurrent relaxation only depends on a certain term $\Delta H$ in the Hamiltonian that does not commute with the total supercurrent operator.

Our first results concern systems in which parity and time reversal symmetries are unbroken (or, at least, where the effects of symmetry breaking due to e.g. a magnetic field can be neglected). We show that fluctuating superconductivity leads to the conductivity
\be
\sigma = \frac{\rho_s}{m^2} \frac{1}{- i \omega + \Omega} + \sigma_0 \,.
\ee
The superfluid relaxation rate $\Omega$ is given by (\ref{eq:key}) below, $\sigma_0$ is the contribution of the normal fluid component to the conductivity and $\rho_s$ and $m$ are susceptibilities that will be defined below. A universal term in the low energy Hamiltonian leading to a nonzero $\Omega$ in the presence of mobile vortices is $\Delta H = \frac{\lambda}{2} \int d^2x \, \rho(x)^2$ -- see equation (\ref{eq:coul}) below. If the vortices are sufficiently large, this interaction leads to
\be
\Omega = \frac{\rho_s}{m^2} \frac{n_f \pi r_v^2}{2\,\sigma_\text{n}} \,.
\ee
This is equation (\ref{eq:OO}) below. Here $n_f$ and $r_v$ are, respectively, the number density and radius of mobile (free) vortices and $\sigma_\text{n}$ is the conductivity of the normal state. In particular, this expression recovers exactly a classic `Bardeen-Stephen' result \cite{BS, halperinnelson} for the d.c. resistivity due to vortices. Our approach embeds that result in a more general, transparent and fully quantum framework.

We proceed to incorporate strong parity and time reversal symmetry breaking. The longitudinal and Hall -- $\sigma$ and $\sigma^H$ --  conductivities are now given by
\be
\sigma^H + i \, \sigma = \frac{\rho_s \, (1 + \r_v^2)}{m^2} \frac{\Omega^H + i \Omega + \omega}{\left(- i \omega + \Omega \right)^2 + \left( \Omega^H \right)^2} + \sigma_0^H + i \sigma_0 \,.
\ee
As above, $\sigma_0$ and $\sigma_0^H$ are the normal fluid component conductivities.
The frequency-dependent response reveals what we will call a `hydrodynamic supercyclotron' mode at
\be
\omega_\star = \pm \Omega^H - i \Omega \,.
\ee
That is to say, the mode oscillates at frequency $\Omega^H$ with decay rate $\Omega$. If $\Omega^H$ is large enough compared with $\Omega$, the peak in the optical conductivity moves away from $\omega=0$. This mode is analogous to the hydrodynamic cyclotron mode in a magnetic field (e.g. \cite{Hartnoll:2007ih}), but supported by a long-lived supercurrent rather than a long-lived momentum. Both $\Omega$ and $\Omega^H$ depend on the supercurrent-relaxing Hamiltonian $\Delta H$. As an example of this physics, we consider a nonlocal interaction $\Delta H = \frac{\lambda'}{2} \int \frac{d^2k}{(2\pi)^2} \frac{\rho_{-k} \left(\nabla \times j \right)_k^z}{k^2} + \text{h.c.} \,.$ As we explain in section \ref{sec:CS}, this interaction is equivalent to coupling the superfluid to an emergent Chern-Simons gauge field. This field creates superfluid vortices whose motion degrades the supercurrent. The supercyclotron mode in this case is found to have
\be
\omega_\star = \frac{\lambda' \rho_s}{m^2} \frac{\pm 1 - \lambda' (\pm \sigma_0^H + i \sigma_0)}{(1-\lambda' \sigma_0^H)^2 + (\lambda' \sigma_0)^2} = \frac{\lambda' \rho_s}{m^2} \frac{1}{\pm 1 - \lambda' (\pm \sigma_0^H - i \sigma_0)} \,.
\ee
This is equation (\ref{eq:scc}) below. The complex frequency of the mode is proportional to the superfluid density $\rho_s$ and the Chern-Simons coupling $\lambda'$. Furthermore, $\Omega$ is proportional to the incoherent conductivity $\sigma_0$. This `incoherent conductivity' quantifies the dissipation
caused by the motion of non-superfluid charged excitations (i.e. the `normal fluid component' -- cf. \cite{Davison:2015taa}). In this example, therefore, dissipation is not caused by the vortex cores.

\subsection{Mechanisms of low temperature dissipation}

Beyond the specific examples of phase-disordering interactions summarized above, the formalism we develop gives a clear perspective on possible quantum mechanisms for supercurrent relaxation as $T \to 0$, and hence for metallic phases in two dimensions. Specifically, we will see that:
\begin{enumerate}
\item Local (short range) charge density interactions can only lead to supercurrent relaxation in the presence of mobile vortices with dissipative cores. Conventional vortices, if they are present at all, are not expected to remain mobile at $T = 0$ \cite{LO, FFH}.

\item Nonlocal charge density interactions can result in supercurrent relaxation due to dissipative processes entirely outside of vortex cores. These dissipative processes will involve the normal fluid component of the system. Therefore, if a normal fluid component survives to $T=0$ then, in principle, so can supercurrent relaxation.

\item The existence of a normal fluid component in itself is not sufficient to relax the supercurrent. Typically the supercurrent simply short circuits the normal fluid. Gapless excitations that might mediate long range nonlocal interactions are also not guaranteed to relax the supercurrent, they will simply themselves be part of the normal fluid. To disorder the phase and relax the supercurrent, the nonlocal interaction needs to have specific properties. The Chern-Simons interaction we consider below is an example of an interaction that does the job. By creating vortices through flux attachment, it ties the dynamics of the charge density (including the normal component) to phase-fluctuation physics.

\end{enumerate}

\section{Superfluid hydrodynamics of incoherent metals}

There are two important sources of infinite dc conductivities in systems with a nonzero charge density. Firstly, in a superfluid phase, an infinite conductivity follows from conservation of the supercurrent operator $J_\phi$. Secondly, in a translationally invariant system, an infinite conductivity follows from conservation of the total momentum $P$. Both conservation laws must be broken to obtain a finite conductivity.

Relaxation of momentum can be achieved by disorder, umklapp scattering or coupling to a momentum-non-conserving bath (e.g. phonons away from the phonon drag regime).
Weak momentum relaxation in the normal, non-superconducting, state results in the metal entering a hydrodynamic regime \cite{Hartnoll:2007ih, spivak1, stevefluid, Lucas:2015sya}. Hydrodynamic metals exhibit unconventional physics that is currently of considerable experimental interest \cite{geimfluid,subirfluid,andyfluid}.

The opposite limit of very strong momentum relaxation (but without localization) leads to `incoherent metals' \cite{Hartnoll:2014lpa}. It is possible that many of the most interesting strongly correlated systems are in this class: for instance, many are close to localized phases and exhibit very broad Drude peaks, if they have Drude peaks at all \cite{Hartnoll:2014lpa}. The essence of an incoherent metal is that there is no advective transport and hence the only hydrodynamic variables
are fluctuations of the charge $\rho$ and energy $\epsilon$ densities. In particular, the local velocity $u$ does not appear as a hydrodynamic variable and hence there are no sound modes \cite{Davison:2014lua}. The densities obey the conservation equations
\be
\frac{\pa \epsilon}{\pa t} + \nabla \cdot j^E = 0 \,, \qquad \frac{\pa \rho}{\pa t} + \nabla \cdot j = 0 \,.
\ee
Here $j$ and $j^E$ are the electric and energy current, respectively. The Green's functions for the conserved densities are obtained using constitutive relations. These capture dissipative physics in a gradient expansion, as we will recall shortly.

Our object of study here is an incoherent metal that attempts but fails to become superconducting. In a superfluid phase an additional hydrodynamic variable appears: the superfluid velocity \cite{chaikin}
\be\label{eq:uphi}
u_\phi = \frac{1}{m} \nabla \phi \,.
\ee
Here $\phi$ is a long wavelength perturbation of the superfluid phase. The constant $m$ is a mass scale, that we discuss futher later (essentially, $m^{-1}$ will be the susceptibility $\chi_{j u_\phi }$, defined below). The static susceptibility\footnote{The superfluid density appears in the free energy density as $f = \cdots + \half \rho_s u_\phi^2$. The free energy is itself a function of $u_\phi$ and must be Legendre transformed to obtain the thermodynamic potential for the superfluid source $h_\phi = \pa f/\pa u_\phi = \rho_s u_\phi$. This potential is then $g = \cdots - \half h_\phi^2/\rho_s$. The susceptibility (\ref{eq:chiphi}) then follows from $\chi = - \pa^2 g/\pa h_\phi^2$.} of $u_\phi$ defines the superfluid density $\rho_s$:
\be\label{eq:chiphi}
\chi_{u_\phi^i u_\phi^j} = \frac{1}{\rho_s} \delta^{ij} \,.
\ee
We will study the effects of quantum phase fluctuations that relax the superfluid velocity, and hence frustrate the attempt to become superconducting, but leave $\rho_s$ finite. Superfluid hydrodynamics will remain useful if $u_\phi$ relaxes sufficiently slowly. We will work in this limit, in which we will be able to get a theoretical handle on the problem. In this limit there is a sharp Drude-like peak in the optical conductivity with width given by the supercurrent decay rate $\Omega$. This decay rate, and hence the d.c. conductivities, can be obtained using the memory matrix formalism \cite{forster}. The power of this approach is that it packages all microscopic details into a single quantity, $\Omega$. The important input -- beyond simple hydrodynamics --  is a term $\Delta H$ in the Hamiltonian responsible for phase relaxation (i.e. such that $[\Delta H, J_\phi]\neq 0$). An elegant memory matrix discussion of phase fluctuations in one spatial dimension exists \cite{andrei}. We will be considering the case of two spatial dimensions. An important part of this work will be the identification of interesting and natural terms $\Delta H$. First, however, we must describe the hydrodynamic framework with a conserved supercurrent.

The memory matrix has been successfully used in recent years to obtain the d.c. conductivity of hydrodynamic metals with slow momentum relaxation \cite{Hartnoll:2007ih, Hartnoll:2008hs, Hartnoll:2012rj, Hartnoll:2014gba, Lucas:2015pxa}. By focusing on incoherent metals, we can concentrate on cases where the dc conductivity is determined solely by the physics relaxing the supercurrent. In a separate work we will consider the interplay of both momentum and supercurrent relaxation \cite{todo}.

To obtain the equations of motion for the hydrodynamic variables $\rho, \epsilon, u_\phi$ we need to write down the constitutive relations for the two conserved currents as well as the `Josephson relation' for the phase \cite{chaikin}. The physics is most transparent if we swap the energy density and current for the entropy density $s$ and heat current $j^Q$:
\bea
d \epsilon & = & T d s + \mu d \rho + \rho_s u_\phi \cdot d u_\phi \,, \label{eq:deltae} \\
j^Q & = & j^E - \mu j \,.
\eea
We are considering linear response about a state with no supercurrent, so the last term in (\ref{eq:deltae}) will be subleading for most purposes. To first order in a derivative expansion the constitutive and Josephson relations are
\bea
j -  \frac{\rho_s}{m} u_\phi  & =  &- \a_1 \nabla s - \a_2 \nabla \rho + \cdots  \,, \label{eq:jj} \\
{\textstyle \frac{1}{T}}j^Q & = & - \b_1 \nabla s - \b_2 \nabla \rho + \cdots \,, \label{eq:jq} \\
\pa_t \phi & = & - \mu + \xi \rho_s \nabla \cdot u_\phi + \cdots \,. \label{eq:jose}
\eea
There are five dissipative transport coefficients $\a_1,\a_2,\b_1,\b_2,\xi$. 
The superfluid velocity (\ref{eq:uphi}) should be counted at zeroth order in the derivative expansion. Therefore, it is convenient to take the gradient of the Josephson relation and write
\be\label{eq:udot}
m \, \pa_t u_\phi = - \nabla \mu + \xi \, \rho_s \nabla \, \nabla \cdot u_\phi + \cdots \,.
\ee
The non-dissipative term that we have placed on the left hand side of (\ref{eq:jj}) is fixed by
absence of entropy production to leading order in derivatives (i.e. set $\dot s + \nabla \cdot (j^Q/T) = 0$, using (\ref{eq:deltae}) to calculate time derivatives, including the last term, and the conservation laws).

The above equations assume that two dimensional parity is unbroken. In section \ref{sec:noP} below
we will describe the case with broken parity. The parity broken case is experimentally relevant due to the presence of magnetic fields in many studies of quantum fluctuating superconductivity.

From the constitutive and Josephson relations combined with the conservation laws, the thermoelectric conductivities can be obtained following Kadanoff and Martin \cite{km}. In practice, it is simpler to use the hydrodynamic equations of motion to eliminate $u_\phi$ and hence write
\be
\left(
\begin{array}{c}
j \\
\frac{1}{T} j^Q
\end{array}
 \right)
=
\left(
\begin{array}{cc}
\sigma & \alpha \\
\alpha & \overline \kappa/T
\end{array}
\right)
\left(
\begin{array}{c}
- \nabla \mu \\
- \nabla T
\end{array}
\right) \,, \label{eq:cmat}
\ee
from which the conductivities immediately follow (with space and time dependence $e^{-i\omega t + i k \cdot x}$, and setting the wavevector $k$ to zero in the matrix of conductivities).
Thus we obtain
\bea
\sigma & = & \frac{\rho_s}{m^2} \frac{i}{\omega} + \sigma_0 \,, \label{eq:s0} \\
\alpha & = & \alpha_0 \,, \label{eq:a0} \\
\overline \kappa & = & \overline \kappa_0 \label{eq:k0} \,.
\eea
The conductivity (\ref{eq:s0}) amounts to a `two-fluid' description in which there is a superfluid and normal contribution to the conductivity. The normal fluid in this case is completely incoherent, with no sound mode due to the absence of a long-lived momentum. The incoherent parts of the above expressions are given by the Einstein relations (c.f. \cite{Hartnoll:2014lpa})
\be\label{eq:incoh}
\left(
\begin{array}{cc}
\sigma_0 & \alpha_0 \\
\alpha_0 & \overline \kappa_0/T
\end{array}
\right) =\left(
\begin{array}{cc}
\a_2 & \alpha_1 \\
\b_2 & \b_1
\end{array}
\right) \cdot \left(
\begin{array}{cc}
\chi_{\rho\rho} & \chi_{\rho s} \\
\chi_{s\rho} & \chi_{ss}
\end{array}
\right) \,.
\ee
The symmetry of the matrix of conductivities (Onsager relation) imposes one constraint on the dissipative transport coefficients $\a_1,\a_2,\b_1,\b_2$. The susceptibilities are of course symmetric so that $\chi_{\r s} = \chi_{s \r}$. Note that $T \chi_{ss} = c_\mu$, the specific heat at constant chemical potential. $\bar \kappa$ is the thermal conductivity at vanishing electric field (closed circuit boundary conditions). As expected, the electrical conductivity diverges as $\omega\to 0$. We have not yet incorporated the effect of phase fluctuations.

The normal modes of the hydrodynamic system above are easily seen to be a pair of `second sound' (although there is no normal sound in this case) modes
\be
\omega(k) = \pm \sqrt{\frac{\rho_s c_\mu}{m^2 T \det \chi}} k - \frac{i}{2} \left(\frac{\rho_s \xi}{m} - \frac{\overline \kappa_0}{c_\mu} + \frac{c_\mu \sigma_0 + \chi_{\r\r} \overline \kappa_0 - 2 T \chi_{s\rho} \alpha_0}{T \det \chi} \right) k^2 \,, \label{eq:sound}
\ee
where $\det \chi = \chi_{\r\r} \chi_{ss} - \chi_{s \r}^2$, and a heat diffusion mode
\be\label{eq:diff}
\omega(k) = - i \frac{\overline \kappa_0}{c_\mu} k^2 \,.
\ee
The thermal diffusivity in this, superfluid, case is therefore proportional to the closed circuit thermal conductivity $\overline \kappa$. In the non-superfluid case, it is the open circuit thermal conductivity $\kappa$ that appears \cite{Hartnoll:2014lpa}. In a metal, the charge-carrying sound modes (\ref{eq:sound}) are of course screened by Coulomb interactions, which give the modes a mass. However, measurements of both dc and optical electrical conductivities measure the current response to the total rather than the external electric field. They are therefore given by the unscreened Green's functions. The expression (\ref{eq:s0}) therefore applies to superconductors as well as superfluids, as do expressions for the dc conductivity below such as (\ref{eq:sdc}).

\section{Vortices and supercurrent relaxation (with parity)}
\label{sec:cou}

\subsection{Superfluid hydrodynamics with vortices}

Supercurrent relaxation is necessarily tied up with vortex physics. It is convenient to consider the system on a spatial torus. A nonzero supercurrent implies a winding of the phase, which can only be relaxed by topological defects. Vortices are defects in the superfluid velocity. While the superfluid velocity is locally a gradient (\ref{eq:uphi}), at a vortex a quantized circulation is present. If $n_v$ is the local density of vortices, then
\be\label{eq:nvdef}
\epsilon^{ij} \nabla^i u_\phi^j = \frac{2 \pi}{m} n_v\,.
\ee
In a continuum description, the incorporation of vortices requires a transverse part in the superfluid velocity. Therefore, for a global description, we must generalize (\ref{eq:uphi}) to
\be\label{eq:phipsi}
u^i_\phi = \frac{1}{m} \Big(\nabla^i \phi + \epsilon^{ij} \nabla^j \psi \Big) \,.
\ee
However, microscopically speaking, $n_v$ comes from coarse-graining over many separated vortex cores. Outside of the vortex cores the vorticity vanishes and hence $\nabla^2 \psi = 0$. This means that locally, outside of vortex cores, $u_\phi$ can be written as the gradient of a phase. The superfluid hydrodynamics we are about to describe takes place outside the vortex cores, even while it depends on the local density $n_v$ of vortices.

Because vortices are topological defects, they can only disappear within low energy dynamics through annihilation with an anti-vortex. Therefore, the local vorticity is conserved and there exists a vortex current $j_v$ satisfying
\be\label{eq:nv}
\frac{\pa n_v}{\pa t} + \nabla \cdot j_v = 0 \,.
\ee
Equation (\ref{eq:nvdef}) means that we can always trade the vortex density $n_v$ for the curl of the superfluid velocity. The superfluid velocity is a zeroth order variable in the gradient expansion and hence the vortex density is first order.

Vortices modify the Josephson relation (\ref{eq:udot}) to
\be\label{eq:jmod}
m \pa_t u^i_\phi = - \nabla^i \mu + 2\pi \epsilon^{ij} j^j_v + \xi \, \rho_s \nabla^i \, \nabla^j u^j_\phi + \cdots \,.
\ee
Taking the curl of this equation, we see that it implies the vorticity conservation law (\ref{eq:nv}).
The new term in the `generalized Josephson relation' above has a direct physical interpretation: a flow of vortices induces a transverse electrostatic potential gradient. In a magnetic field this dynamics underlies, for instance, the vortex Nernst signal \cite{old, ong}. The fact that a vortex current causes a time dependence in the perpendicular phase gradient will ultimately allow relaxation of the supercurrent.

The constitutive relations for the charge and heat currents -- (\ref{eq:jj}) and (\ref{eq:jq}) above -- are not changed by the presence of vortices. We must add a new constitutive relation for the vortex current
\be\label{eq:jvcons}
j^i_v - \frac{m}{2\pi} \Omega \, \epsilon^{ij} u^j_\phi =  - \gamma \nabla^i n_v + \cdots \,.
\ee
The `intrinsic vortex diffusivity' $\gamma$ in (\ref{eq:jvcons}) must be positive. $\gamma$ will not play an important role in our discussion. More important is the $\Omega$ term in (\ref{eq:jvcons}).
The $\Omega$ term is allowed by parity and is the analogue (in our incoherent limit with no conserved momentum) of the superfluid Magnus force. It has some similarity with the term appearing on the left hand side of the electric current constitutive relation (\ref{eq:jj}). As we noted, however, the vortex density and current are already first order in the gradient expansion (unlike the charge density and electric current). $\Omega$ itself must therefore be counted as first order in derivatives, and leads to dissipation. It is required by positivity of entropy production to satisfy $\Omega > 0$. A formula for $\Omega$ will be obtained in the following section \ref{sec:memA}. Indeed, because the $ \Omega \, \epsilon^{ij} u^j_\phi$ term in (\ref{eq:jvcons}) introduces a transverse part into the vortex current, it `breaks' the Josephson equation (\ref{eq:jose}) for the phase to
\be
\pa_t \phi = - \mu - \Omega \, \phi + \cdots \,. \label{eq:broken}
\ee
It follows that a nonzero $\Omega$ is tantamount to saying that $\phi$ is no longer a Goldstone boson. The precise meaning of $\Omega$ will become clearer in the following section \ref{sec:memA}.
It will not, therefore, be a surprise when we find shortly that it gaps out various hydrodynamic modes, leading to a finite dc conductivity. These effects can still be captured within hydrodynamics so long as $\Omega$ is much smaller than the local equilibration rate (presumably set by the temperature and chemical potential). 

Extracting the matrix of conductivities (\ref{eq:cmat}) from the hydrodynamic equations as above,
now gives the Drude-like form for the electrical conductivity
\be\label{eq:s1}
\sigma = \frac{\rho_s}{m^2} \frac{1}{- i \omega + \Omega} + \sigma_0 \,.
\ee
The thermoelectric and thermal conductivities are unchanged from (\ref{eq:a0}) and (\ref{eq:k0}) by the presence of vortices (recall we consider a parity-invariant theory). The dc electrical conductivity is now finite and given by
\be\label{eq:sdc}
\sigma_\text{dc} = \frac{\rho_s}{m^2} \frac{1}{\Omega} + \sigma_0 \,.
\ee
The coefficients $\rho_s$ and $m^2$ here -- we will see later that they are both thermodynamic susceptibilities -- are those in the theory with a nonzero $\Omega$.

While we will mostly focus on charge transport, it is clear from the electrical (\ref{eq:sdc}) and thermal (\ref{eq:k0}) conductivities that the Weidemann-Franz law will be strongly violated in the fluctuating superconductivity regime, with Lorenz ratio
\be\label{eq:Lor1}
L \equiv  \frac{\kappa}{\sigma T} ~\sim \frac{\overline \kappa_0  m^2}{\rho_s} \frac{\Omega}{T} \ll 1 \,.
\ee
Recall that the open circuit thermal conductivity $\kappa = \overline \kappa - \alpha^2 T/\sigma$.

The collective hydrodynamic modes are now seen to be as follows. The thermal diffusion mode (\ref{eq:diff}) is unaffected to leading order at small $\Omega$. A new mode appears which describes the dynamics of the vorticity. This transverse part of the superfluid velocity was previously inert. This mode is a gapped diffusive mode
\be
\omega(k) = - i \Omega - i \gamma k^2 \,.
\ee
The `second sound' modes (\ref{eq:sound}) become one gapped and one ungapped diffusive mode. For small $\Omega$, the gapped mode has the dispersion
\be
\omega(k) = - i \Omega + i \frac{\rho_s}{m^2}\frac{c_\mu}{T \det \chi} \frac{k^2}{\Omega} \,,
\ee
(note that this is the dispersion in the limit in which $k \to 0$ is taken before small $\Omega$, so the mode is causal) while the new gapless diffusive mode has
\be
\omega(k) = - i \frac{\rho_s}{m^2}\frac{c_\mu}{T \det \chi} \frac{k^2}{\Omega} \equiv - i D_\Omega \, k^2 \,.
\ee
The diffusivity $D_\Omega$ of this last mode is the speed of the unrelaxed sound mode (\ref{eq:sound}) squared, divided by the superfluid relaxation rate $\Omega$. A diffusive mode with a large diffusivity, of order $D_\Omega \sim 1/\Omega$, should have been anticipated on general grounds in order for the finite dc conductivity (\ref{eq:sdc}) to obey an Einstein relation. That is (to leading order as $\Omega \to 0$)
\be
\sigma_\text{dc} = \frac{T \det \chi}{c_\mu}\,  D_\Omega \,.
\ee
This expression is very much in the spirit of classic studies of the BKT phase in two dimensional superconductors, such as \cite{halperinnelson}, that obtain the conductivity in terms of a `vortex mobility' proportional to $\Omega$. Note that any `pinning forces' have already been accounted for by working in the incoherent limit with no conserved momentum. Note also that it is $\Omega$ rather than the intrinsic diffusivity $\gamma$ in (\ref{eq:jvcons}) that determines the dominant dissipative motion of the vortices here.

As in our discussion in the previous section of normal modes without vortices, the (now diffusive) charge-carrying mode is gapped by dynamical Coulomb interactions. However, also as above, because the conductivity is defined as the current induced by the total (rather than external) electrical field, optical and dc conductivities are computed from the unscreened Green's functions.
Electromagnetism also alters the long range interactions between vortices. This can be ignored so long as the sample is sufficiently thin \cite{pearl}.

\subsection{Supercurrent relaxation from the memory matrix}
\label{sec:memA}

A formula for $\Omega$ can be obtained using the memory matrix method. 
This method will be useful if the underlying Hamiltonian of the system can be written as
\be
H = H_0 + \vep \Delta H \,,
\ee
such that
\be
[H_0,J_\phi] = 0 \,, \qquad \text{but} \qquad \dot J_\phi = \vep \, i [\Delta H,J_\phi] \neq 0 \,. \label{eq:slow}
\ee
Here $J_\phi$ is the total supercurrent operator, to be defined more precisely below, and $\vep$ is, for the moment, a formal small expansion parameter. The point is that we wish to treat the supercurrent-relaxing physics perturbatively. In the theory with $\vep = 0$ the supercurrent is conserved and hence $\Omega = 0$. The memory matrix formalism will now allow us to obtain a perturbative formula for $\Omega$, which will be of order $\vep^2$.

The electrical conductivity is given by \cite{forster}
\be\label{eq:fullmemory}
\sigma_{JJ}(\omega) = \sum_{CD} \chi_{J C} \left(\frac{1}{- i \omega \chi +  M(\omega) + N}\right)_{CD} \chi_{D J} \,.
\ee
Here the sum runs over both the long lived operator (the total supercurrent $J_\phi$) as well as the external hydrodynamic current (the total electric current $J$). That is, $\{C,D\} \in \{J, J_\phi\}$. The first thing that (\ref{eq:fullmemory}) says is that the weight of the Drude-like peak in the conductivity is given by the $\chi$'s. We will see that these determine the overlap between the various current operators with the supercurrent. The width and location of the peak are determined by the matrices $M$ and $N$.

We proceed to define $\chi,M,N$. We give more general definitions than are needed in this section, for later use. In particular, the formulae quoted hold in the absence of time-reversal invariance.

The static susceptibility of two operators $C$ and $D$ is given in terms of the retarded Green's function as
\bea
\chi_{CD} & \equiv & \frac{1}{2 \pi} \int_{-\infty}^\infty \Big( \text{Im} \, G^R_{CD}(\omega)+\text{Im} \, G^R_{DC}(\omega) \Big)  \frac{d\omega}{\omega}  \label{eq:chi}  \\
& = & \frac{1}{2 \pi} \int_{-\infty}^\infty \Big( \text{Im} \, G^R_{CD}(\omega,B)+ \eta_C \eta_D\text{Im} \, G^R_{CD}(\omega,-B) \Big)  \frac{d\omega}{\omega}\,.
\eea
The second line here reminds us that in the absence of time reversal, e.g. in the presence of a magnetic field $B$, the two terms in the integrand are not always equal. $\eta_{C/D}$ are $\pm 1$ depending on whether the operators are even or odd under time reversal. The susceptibilities as defined above are equal to the thermodynamic susceptibilities \cite{forster}. They are symmetric, even without time reversal symmetry: $\chi_{CD} = \chi_{DC}$. Let us write
\be \label{eq:chi_memory}
\chi_{JJ_\phi} = \frac{1}{m} \,, \qquad \chi_{J_\phi J_\phi} = \frac{1}{\rho_s} \,.
\ee
These should be taken to be the definitions of $m$ and $\rho_s$ in this approach.

The matrix $N$ is given by (note the time derivative on one of the operators)
\be
N_{CD} \equiv \chi_{C \dot D} = - \chi_{\dot C D} \,. \label{eq:NN}
\ee
In the present, time-reversal invariant case, $N = 0$. All $\{C,D\} \in \{J, J_\phi\}$ are odd under time reversal. Therefore the derivative $\dot C$ has the opposite time reversal transformation to $D$ and hence their overlap in a time-reversal invariant state is zero.

The memory matrix $M$ is given by
\be
M_{CD}(\omega) \equiv \frac{i}{T} \bigg(\dot C \; \bigg| \; {\mathcal Q} \frac{1}{\omega - {\mathcal Q} L {\mathcal Q}} \mathcal Q \; \bigg| \; \dot D \bigg) \,. \label{eq:MMM}
\ee
We will not need to define the inner product of operators $(A|B)$ here, see \cite{forster}. The quantum Liouville operator $L = [H,\bigcdot \,]$. All we need to know about the projection operator ${\mathcal Q}$ is that, with respect to this inner product, it projects onto the space of operators orthogonal to the set $\{J,J_\phi\}$ that are being summed over in the basic expression (\ref{eq:fullmemory}). In a time-reversal invariant state, it can be shown that, working to leading nontrivial order in $\vep$, the projection operators have no effect, and one can set ${\mathcal Q} = 1$. The argument leading to this conclusion is not as simple as has been claimed in past works. We give a correct argument in appendix \ref{sec:Qe}. With the projectors set to unity, one has \cite{forster}
\be
M_{CD}(\omega) = \frac{1}{i \pi} \int_{-\infty}^\infty \frac{\text{Im} \, G^R_{\dot C \dot D}(\omega') \, d\omega'}{\w' (\w' - \w)} \,.
\ee
The integral is regularized by taking $\omega$ to have a small positive imaginary part. We will only need the result for small frequencies. Standard Kubo-formula type manipulations give
\be\label{eq:mem}
M_{CD}(0) = \lim_{\omega \to 0} \frac{\text{Im} \, G^R_{\dot C \dot D}(\omega)}{\omega} \,.
\ee
With time reversal invariance, the memory matrix is symmetric: $M_{CD} = M_{DC}$.

One more quantity that will be useful to introduce is the `incoherent susceptibility' \cite{Davison:2015taa}
\be
\chi_{JJ}^\text{inc.} = \chi_{JJ} - \frac{\rho_s}{m^2} \,.
\ee
This is the susceptibility of the incoherent current operator $J^\text{inc.} \equiv J - \frac{\rho_s}{m} J_\phi$.

The crucial point is that the memory matrix is proportional to the time derivative of operators. Components of this matrix are therefore small if the corresponding operators are long-lived. In particular, $M_{J_\phi J_\phi} \sim \vep^2$ follows immediately from the definition (\ref{eq:mem}) together with the fact that $J_\phi$ is long-lived according to (\ref{eq:slow}). This observation allows us to consider the following scaling limit of the full expression (\ref{eq:fullmemory}) as $\vep \to 0$:
\be
\omega \sim M_{J_\phi J_\phi} \sim \vep^2 \,, \qquad M_{J J_\phi} \sim \vep^2 \,, \qquad \chi_{JJ}^\text{inc.} \sim \frac{1}{\vep} \,. \label{eq:limit}
\ee
All other quantities remain order one. Two comments are in order. Firstly, one might have anticipated $M_{J J_\phi} \sim \vep$, because $J$ is not a long-lived operator. However, in appendix \ref{sec:Qe} we show that in some generality, in fact $M_{J J_\phi} \sim \vep^2$. An analogous fact was noted in \cite{Hartnoll:2012rj}. Secondly, the final assumption that the incoherent susceptibility be large is not essential, but is needed in order for the incoherent contribution to the conductivity -- $\sigma_0$ in (\ref{eq:s1}) -- to appear at the same order in $\vep$ as the fluctuating superfluid contribution. This limit thereby avoids ambiguities in the incoherent contribution due to spectral weight transfer from the Drude peak \cite{Davison:2015bea, Blake:2015epa}. A similar, but slightly different, limit was considered in \cite{Lucas:2015pxa}. 

In the scaling limit (\ref{eq:limit}) the memory matrix expression (\ref{eq:fullmemory}) recovers the hydrodynamic result (\ref{eq:s1})
\be
\sigma = \frac{\rho_s}{m^2} \frac{1}{- i \omega + \Omega} + \sigma_0 \,,
\ee
but now with the microscopic formulae for $\Omega$ and $\sigma_0$:
\be
\Omega = \left. \vep^2 \rho_s \lim_{\omega \to 0} \frac{\text{Im} \, G^R_{i [\Delta H,J^x_\phi] \, i [\Delta H,J^x_\phi]}(\omega)}{\omega} \right|_{\vep = 0} \,,\label{eq:key}
\ee
and
\be
\sigma_0 = \frac{\left(\chi_{JJ}^\text{inc.}\right)^2}{M_{JJ}(0)} \,.
\ee
The thermoelectric and thermal conductivities, (\ref{eq:a0}) and (\ref{eq:k0}), are also reproduced.
The retarded Green's function on the right hand side of (\ref{eq:key}) is to be evaluated in the unperturbed theory with a conserved supercurrent operator. This type of formula for $\Omega$ goes back to the seminal work \cite{GW}. See also \cite{andrei2} for a helpful discussion. The central and useful fact is that $\Omega$ depends on a correlation function of $\dot J_\phi$, the time derivative of a long lived operator, as opposed to the conductivity itself which is just given by the correlation function of $J$. For recent uses of this type of formula in the case of slow momentum relaxation, see e.g. 
\cite{Hartnoll:2008hs,Hartnoll:2012rj,Hartnoll:2014gba,Lucas:2015pxa}. Supercurrent relaxation in one spatial dimension has been described in this language in \cite{andrei}.

To evaluate the key formula (\ref{eq:key}) we must of course specify $\Delta H$. We first discuss $J_\phi$. The supercurrent density operator is defined outside of the vortex cores to be the gradient of the local phase of the order parameter
\be\label{eq:jphilocal}
j_\phi = \frac{1}{m} \nabla \phi \,.
\ee
(Note that $j_\phi$ refers to the operator, while above $u_\phi = \langle j_\phi\rangle$).
As discussed below (\ref{eq:phipsi}) above, this is not a globally defined operator in the presence of the vortices. However, it is well defined outside of the vortex cores, allowing for windings of the phase, in which we identify $\phi \sim \phi + 2 \pi$. The total supercurrent operator which is to be relaxed is then
(we will be interested in the case of two spatial dimensions)
\be\label{eq:jphi}
J_\phi = \frac{1}{m} \int_{T^2 \setminus \{\text{vortex cores}\}} d^2x \nabla \phi \,.
\ee
Here we have placed the theory on a spatial torus. This is the standard way to describe a supercurrent, which becomes the winding of the phase around the torus. The expressions
(\ref{eq:jphilocal}) and (\ref{eq:jphi}) are also the definition of the mass scale $m$ (we will see below that this is the same as defining $1/m$ to be the susceptibility $\chi_{JJ_\phi}$).

We have just seen that $J_\phi$ is defined in terms of the local phase of the superfluid order parameter. Now, the momentum conjugate to this phase is the charge density $\rho$.\footnote{Using the Josephson relation (\ref{eq:jose}), the conjugate momentum $\pi_\phi = \frac{\pa f}{\pa \dot \phi} = - \frac{\pa f}{\pa \mu} = \rho$.}  That is, at equal times,
\be\label{eq:can}
[\phi(x), \rho(y)] = i \delta(x-y) \,.
\ee
Therefore, it is natural to build operators that do not commute with the supercurrent (\ref{eq:jphi}) out of the charge density. Evaluating the commutator can be subtle, however, because the supercurrent operator (\ref{eq:jphi}) is the integral of a total derivative. The operator is not zero because $\phi$ admits a shift symmetry and can therefore have winding when placed on a spatial torus. We will need to find interactions that can have a nontrivial commutator with the total derivative operator $J_\phi$. In the first case we consider, the vortex cores, which define boundaries of the region where the phase is defined, will be crucial.

Before moving on to describe the first example of an interaction $\Delta H$, we should pause to summarize where we are. Our entry point was the assumption that the system could approximately be described by superfluid hydrodynamics. By considering `incoherent' hydrodynamics, without a long lived momentum density, we have essentially integrated out the effects of disorder from the start. The incoherent (`diffusive') part of the conductivities (\ref{eq:incoh}) could be metallic or insulating; it does not matter at this point, because our assumption is that the dc conductivity (\ref{eq:sdc}) is dominated by the long lived supercurrent.
It should be clear from the expression for the supercurrent relaxation rate (\ref{eq:key}) that we are considering quantum phase-disordering processes. We are not considering `paraconductivity' physics in which the superfluid itself fluctuates into existence above the critical temperature. Similarly, `amplitude fluctuations' of the order parameter below the critical temperature do not relax the supercurrent and their effects are implicitly contained in the $\chi,M,N$ matrices.

\subsection{Supercurrent relaxation from short range charge density interactions}
\label{sec:BKT}

In the search for interactions $\Delta H$ that weakly degrade the supercurrent, we will pursue an effective field theory approach. That is, we will write down simple interactions that are consistent with the symmetries of the long wavelength superfluid hydrodynamics. The coupling constants of these interactions will be undetermined numbers, but are generically expected to be nonzero. For this approach to be consistent, we must check that indeed the effects of these interactions can be captured perturbatively when the coupling constants are small. They will then describe a theoretically controlled perturbation of the low energy superfluid hydrodynamics. In this work we will focus on two natural and interesting interactions. Given the plethora of experimental systems of interest discussed in section \ref{sec:exp} above, it will be important to search for further mechanisms in the future. Within the effective field theoretic framework developed in this paper, it may be possible to perform this search systematically. See the final discussion section \ref{sec:discuss}, where we also discuss percolation scenarios.

The most universal term (in the sense of being least sensitive to short distance details) we have found is perhaps the Chern-Simons interaction described below in section \ref{sec:CS}. That term, however, breaks parity and therefore requires a more complicated hydrodynamic description, that we develop below. In this section we will consider a particular $\Delta H$ that preserves parity and time reversal and is therefore present even in the absence of a magnetic field. One upshot of our discussion here will be an elegant rederivation of established results for the resistivity due to flux flow in a BKT phase (e.g. \cite{halperinnelson,BS}). This gives -- among other things --  a sanity check for our approach.

A number of previous works have studied microscopic models for the competition between phase coherence and Coulomb interactions -- in particular with a view to accessing the low temperature quantum regime, e.g. \cite{t1,t2,t2bis,t2ter,t3,t3bis,t4,t5,inui,emery,t6}. Our approach here is a bit different. We are looking for interactions in an effective low energy Hamiltonian that can consistently be treated as small perturbations of a superfluid state. It is not a priori obvious that such interactions exist. Consider, then, the following short range density-density interaction.
\be\label{eq:coul}
\Delta H =  \frac{\lambda}{2} \int d^2x \, \rho(x)^2 \,.
\ee
This term will typically be present in the low energy description, as often $\rho^2$ is just the kinetic term for the phase, because $\rho = \pi_\phi \sim \dot \phi$. See e.g. \cite{t1,t2bis,t2ter}. Such a term drives the fluctuation dynamics in the numerical study \cite{subiroleg}. In some microscopic models, for instance those involving Josephson junction arrays \cite{PWA,t2}, the kinetic term that initially appears is instead $\int d^2x \,(\nabla \rho)^2$. However, such a term in the Hamiltonian will generate the more relevant `on-site' or `self-charging' term (\ref{eq:coul}) under renormalization group flow. In fact, assuming that charge interactions are local in the effective theory, $\lambda$ is just the inverse of the charge susceptibility $\lambda = \chi_{\rho \rho}^{-1}$.
This follows from using the linearized relation $\delta \rho = \chi_{\rho\rho} \, \delta \mu$ in the energy density of (\ref{eq:coul}) and comparing with the expected free energy density $f = \cdots + \frac{1}{2} \chi_{\rho\rho}\, (\delta \mu)^2$.

Using the canonical commutation relation (\ref{eq:can}) one straightforwardly obtains
\bea\label{eq:H1J}
\dot J_\phi = i [\Delta H,J_\phi] & = & \frac{\lambda}{m} \int_{T^2 \setminus \{\text{vortex cores}\}} d^2x \, \nabla \rho(x) \\
& =  & - \frac{\lambda}{m} \int_{\{\text{vortex cores}\}} d^2x \, \nabla \rho(x) \,. \label{eq:sec}
\eea
To obtain the second line we use the fact that the charge density $\rho$ is a single valued operator over the whole spatial torus. Therefore $\int_{T^2} d^2x \, \nabla \rho(x) = 0$. We immediately learn from (\ref{eq:sec}) that the superfluid relaxation is going to depend only on the normal state dynamics of the vortex interior. This conclusion will still hold if we replace (\ref{eq:coul}) with any local charge density interactions. It is an intuitively physically reasonable fact: heat is generated as the vortices are pushed around, but because the exterior superfluid is non-dissipative, the rate at which this heat is generated depends on the internal degrees of freedom of the vortex.\footnote{Vortices that have been pinned by disorder or by freezing into a lattice do not contribute to superfluid relaxation in (\ref{eq:sec}). We can think of them as corresponding to regions where $m=\infty$.}  Furthermore, equations (\ref{eq:H1J}) and (\ref{eq:sec}) show transparently that, while the vortex core dynamics determines the superfluid relaxation, dissipation will occur equally inside and outside of the cores, c.f. \cite{BS, tinkh}.

Inserting the result for the commutator (\ref{eq:sec}) into the expression for the superfluid relaxation rate (\ref{eq:key}), and dropping the formal expansion parameter $\varepsilon$, we obtain
\be\label{eq:Omega1}
\Omega =
\frac{\lambda^2 \rho_s}{m^2} \; n_f \int_\text{core} d^2x \int_\text{core} d^2y \int \frac{d^2k}{(2\pi)^2} e^{i k \cdot (x-y)} \frac{k^2}{2} \lim_{\omega \to 0} \frac{\text{Im} \, G^R_{\rho \, \rho}(\omega,k)}{\omega} \,.
\ee
Here we have made the plausible approximation that the only charge density correlations are within a single given vortex. Thus we limit the integration to a single vortex and multiply by the prefactor $n_f$, which is the density of free vortices (that is, the density of all vortices with any sign for the vorticity -- this quantity does not break parity). We have also used rotational invariance to replace $k_x^2 \to \frac{1}{2} k^2$ inside the integral. The above formula should be valid so long as $\Omega \ll k_B T$. More precisely, $\Omega$ should be much smaller than typical non-hydrodynamic relaxation rates for current density excitations. Factors that help the validity of the computation include a small coupling $\lambda$ and a small superfluid density $\rho_s$.

If the vortex core size is set by microscopic scales, then the expression (\ref{eq:Omega1}) is as far as we can go without a complete microscopic theory. In that case, the superfluid relaxation rate is set by non-universal short distance physics.\footnote{In fact, in two circumstances the low energy spectral weight $\lim_{\omega \to 0} \text{Im} \, G^R_{\rho \, \rho}(\omega,k)/\omega$, appearing in (\ref{eq:Omega1}), is a universal quantity even at microscopic wavevector $k$ \cite{Hartnoll:2012rj}. The first is with Fermi surface kinematics and $k \lesssim 2 k_F$. The second is with (semi-)local quantum criticality, with dynamical critical exponent $z=\infty$. In these cases an interesting universal $\Omega$ -- distinct from the Bardeen-Stephen formula that we rederive shortly in (\ref{eq:OO}) -- can be obtained even from microscopic vortices. This physics will be explored elsewhere.\label{foot:BSBS}} We can do better, however, if the vortex core is sufficiently large that the interior can be treated as the normal state in thermal equilibrium. In this case, neglecting thermoelectric effects (cf. \cite{Hartnoll:2014lpa}), the low energy density-density correlation function will have the diffusive form
\be\label{eq:didi}
G^R_{\rho \, \rho}(\omega,k) = \frac{k^2 D \chi_{\r\r}}{- i \omega + D k^2} \,.
\ee
The normal state charge diffusivity $D$ here is related to the normal state conductivity by the Einstein relation $\sigma_\text{n} = D \chi_{\r\r}$. The susceptibility $\chi_{\r\r}$ is also now that of the normal state. With the Green's function (\ref{eq:didi})
we easily obtain from (\ref{eq:Omega1}) that
\be\label{eq:OO}
\Omega = \frac{\rho_s}{m^2} \frac{n_f \pi r_v^2}{2\sigma_\text{n}} \,.
\ee
Here $r_v$ is the radius of the vortex. We have used the fact, noted above, that $\lambda = \chi_{\rho \rho}^{-1}$. The susceptibility is position-dependent in the presence of vortices (normal state inside, superfluid state outside). All of the derivations above go through in the presence of a spatially dependent coupling $\lambda$. Inside the vortex, the susceptibilities of course take the normal state values. This discussion also goes through for sufficiently large Josephson vortices, because the core in the Josephson barrier will admit a diffusive mode.

The dc conductivity is now given by (\ref{eq:sdc}). In order to facilitate comparison with past work, we will use the fact that the vortex radius is approximately equal to the Ginzburg-Landau correlation length, $r_v \approx \xi_\text{GL}$, in regimes where that description is applicable (see e.g. \cite{tinkh}). We will furthermore restore, in the following formula only, factors of the charge $e_*$ of the condensate that we will otherwise be setting to unity throughout. In particular, $e_*$ appears multiplying the supercurrent operator (\ref{eq:jphilocal}). Thus we obtain 
\be\label{eq:sdcsn}
\sigma_\text{dc} =\frac{2 \sigma_\text{n}}{\pi  n_f \, \xi_\text{GL}^2} \frac{e^2}{e_*^2} \,.
\ee
The result (\ref{eq:sdcsn}) agrees exactly -- setting $e_* = 2e$ -- with that given in, for instance, equation (32a) of \cite{halperinnelson}. The fully quantum derivation given above shows how this result is ultimately connected to $\Delta \rho \Delta \phi \gtrsim \hbar$ physics in a rather universal way, through the effective coupling (\ref{eq:coul}). The classical nature of the Bardeen-Stephen result has been recovered in taking the diffusive form (\ref{eq:didi}) for the charge density correlations. Our treatment is valid away from this limit. We noted one circumstance where a more general formulation may be useful in footnote \ref{foot:BSBS}. In the discussion section \ref{sec:discuss} below we furthermore note that more general local interactions than (\ref{eq:coul}) can lead to different, quantum, formulae for the rate of dissipation in the vortex core.

According to equation (\ref{eq:sdcsn}), the phase-disordering interaction (\ref{eq:coul}) can lead to a finite and nonzero dc conductivity at $T \to 0$ only if (i) there is a density $n_f$ of mobile vortices and (ii) the residual normal state resistivity $\sigma_\text{n}$ is finite and nonzero. While the normal state is insulating in two dimensions, zero conductivity is only realized at exponentially large distance scales, greater than the size of the vortex cores. Whether vortices proliferate or not at a given temperature involves BKT-type dynamics beyond the hydrodynamic approach taken here \cite{bkt}. The presence of vortices is a topological fact assumed in our argument above (we do not have a general formula for $n_f$). 
The expression (\ref{eq:OO}) can also be applied in sufficiently weak magnetic fields, where vortices will certainly be present. The question is then whether or not these vortices are mobile. Vortices are not expected to be mobile at $T=0$, even with a magnetic field \cite{LO, FFH}.

In the following section we will consider a parity-violating interaction that leads to a relaxation rate that is not determined entirely by vortex core dynamics. This is possible for nonlocal interactions $\Delta H$. Whereas any local interaction of the charge density will lead to a dissipation rate given by formulae analogous to (\ref{eq:Omega1}) -- involving an integration over the vortex core -- nonlocal interactions can `undo' the total derivative in the supercurrent (\ref{eq:jphi}).
The objective is to find a nonlocal interaction that leads to a finite and nonzero $\Omega$. An example of a (parity invariant) term that does not work is an unscreened Coulomb interaction in the effective low energy theory:
\be
\Delta H = \frac{\lambda}{2} \int \frac{d^2k}{(2\pi)^2} \frac{\rho_{-k} \rho_k}{k^2} \,.
\ee
One can show that this term leads to a relaxation rate $\Omega$ that depends upon all of space, not just the vortex cores. However, the expression for $\Omega$ diverges upon taking the $\omega \to 0$ limit in (\ref{eq:key}). This does not mean that unscreened Coulomb interactions necessarily destroy superfluidity, just that their effects cannot be computed in the perturbative memory matrix approach. Coulomb interactions are certainly expected to be screened in the low energy effective theory.

\section{Parity-violating superfluid hydrodynamics with vortices}
\label{sec:noP}

Magnetic fields play a central role in many of the experimental systems exhibiting quantum phase fluctuations, as we described in the introduction. In these cases, parity and time reversal symmetries are both broken, while their composition $\sf PT$ is preserved. This leads to the possibility of additional supercurrent-relaxing terms in the effective Hamiltonian and also modifies the structure of the underlying superfluid hydrodynamics. 

\subsection{Hydrodynamic conductivities without parity}

In the absence of parity, new terms are allowed in the hydrodynamic constitutive relations. Firstly, the  vortex current can now be written as
\be\label{eq:jvcons2}
j^i_v - \frac{m}{2 \pi} \Omega \epsilon^{ij} u^j_\phi - \frac{m}{2 \pi} \Omega^H u^i_\phi = - \frac{s_v}{2\pi} \nabla^i T - \frac{\rho_v}{2\pi} \nabla^i \mu - \gamma \nabla^i n_v + \cdots \,.
\ee
The new terms -- relative to (\ref{eq:jvcons}) -- on the right hand side express the fact that thermal and chemical potential gradients will drive a flow of vortices. The coefficients $s_v$ and $\rho_v$ are determined by the entropy and charge at the vortex core relative to the superfluid. They are also proportional to the net vorticity and therefore can only be present if parity is broken. The coefficients $s_v$ and $\rho_v$ are properties of the hydrodynamics at zeroth order in derivatives (because $j_v$ itself is first order, see (\ref{eq:jmod}) above). The factors of $2\pi$ are to clean up formulae below. On the left hand side of the equation we have included an additional `force' term that is allowed once parity is broken.

We will obtain general expressions for $\Omega$ and $\Omega^H$ below, using memory matrix techniques. In the following section \ref{sec:CS} we will develop an explicit example. There is something of a choice in (\ref{eq:jvcons2}) to parametrize the breaking of the Josephson relation -- in the sense of equation (\ref{eq:broken}) -- via the two quantities $\Omega$ and $\Omega_H$. This choice will be seen to capture the physics correctly, although the concrete model of section \ref{sec:CS} works slightly differently due to a long-range superfluid-degrading interaction.

The constitutive relations for the charge and heat currents are also modified to allow for parity-violating terms. These include new non-dissipative terms that are obtained as described below equation (\ref{eq:udot}) above. We have
\bea
j^i -  \frac{\rho_s}{m} \left(\delta^{ij} + \rho_v \epsilon^{ij} \right) u^j_\phi  & =  &- \hat \a^{ij}_1 \nabla^j s - \hat \a^{ij}_2 \nabla^j \rho  + \cdots  \,, \label{eq:jjj} \\
{\textstyle \frac{1}{T}}j^{Q\,i} - \frac{\rho_s}{m} s_v \epsilon^{ij} u^j_\phi  & = & - \hat \b^{ij}_1 \nabla^j s - \hat \b^{ij}_2 \nabla^j \rho + \cdots \,. \label{eq:qqq}
\eea
The terms proportional to $\rho_v$ and $s_v$ are fixed by $\sf PT$ symmetry (Onsager relation). Here the hats indicate that the quantity is a matrix, so that
\be
\hat \a^{ij}_a = \a_a \delta^{ij} + \a_a^H \epsilon^{ij} \,, \qquad \hat \b^{ij}_a = \b_a \delta^{ij} + \b_a^H \epsilon^{ij} \,,
\ee
with $a=1,2$. We noted below (\ref{eq:nv}) above the vortex density itself is already first order in the hydrodynamic expansion. It follows that
gradients of the vortex density are subleading compared to other density gradients and so we have not included them in the constitutive relations (\ref{eq:jjj}) and (\ref{eq:qqq}).

The full set of equations to solve is then the above three constitutive relations, as well as the conservation laws and Josephson relation (these are not changed from the parity-invariant case above). By manipulating these equations we can obtain the longitudinal conductivities via the procedure described around (\ref{eq:cmat}) above. The answers are
\bea
\sigma & = & \frac{\rho_s}{m^2} \frac{(1-\rho_v^2)(-i\omega + \Omega) + 2 \rho_v \Omega^H}{(-i\omega + \Omega)^2 + (\Omega^H)^2}  + \sigma_0 \,, \label{eq:ss} \\
\alpha & = &  -\frac{\rho_s}{m^2} \frac{s_v\rho_v(-i\omega + \Omega) + s_v\Omega^H}{(-i\omega + \Omega)^2 + (\Omega^H)^2} + \alpha_0 \,, \\
\overline \kappa & = & -\frac{\rho_s T}{m^2} \frac{s_v^2(-i\omega + \Omega)}{(-i\omega + \Omega)^2 + (\Omega^H)^2} + \overline \kappa_0 \,. \label{eq:kap}
\eea
The incoherent parts $\sigma_0,\alpha_0, \overline \kappa_0$ are defined as in equation (\ref{eq:incoh}) above. 

The Hall conductivities are similarly obtained as
\bea
\sigma^H & \equiv & \sigma_{xy} \; = \frac{\rho_s}{m^2} \frac{(1-\rho_v^2)\Omega^H + 2 \rho_v (-i\omega+\Omega)}{(-i\omega + \Omega)^2 + (\Omega^H)^2} + \sigma_0^H \,, \label{eq:shsh} \\
\alpha^H_\pm & \equiv &  \alpha_{xy} \; = \frac{\rho_s}{m^2} \frac{s_v(-i\omega + \Omega) - s_v\rho_v\Omega^H}{(-i\omega + \Omega)^2 + (\Omega^H)^2} + \alpha_0^H \,, \\
\overline \kappa^H & \equiv & \overline \kappa_{xy} \; =  -\frac{\rho_s T}{m^2} \frac{s_v^2\Omega^H}{(-i\omega + \Omega)^2 + (\Omega^H)^2} + \overline \kappa_0^H \,. \label{eq:ll}
\eea
The `incoherent' terms in the above expression again correspond to the finite Hall conductivities ($\sigma^H_0,\alpha^H_0, \overline \kappa^H_0$) in the theory without superfluid relaxation.

From (\ref{eq:ss}) and (\ref{eq:shsh}), the dc electrical conductivity can be compactly written in the complexified form
\be \label{eq:dc_cond_PT}
	\sigma_{\rm dc} + i\sigma_{\rm dc}^H 
		= \frac{\rho_s}{m^2} \frac{(1+i\rho_v)^2}{\Omega-i\Omega_H} + \sigma_0 + i\sigma_0^H \, .
\ee
It is also instructive to obtain the fluctuating superfluid contribution to the Nernst signal
\be\label{eq:Nernst}
e_N \equiv \left(\hat \rho \, \hat \alpha \right)_{xy} =  \frac{s_v}{1 + \rho_v^2} \,.
\ee
Here hats indicate that we consider the matrix of conductivities. The final expression is independent of $\Omega$ and $\Omega^H$, and directly relates the Nernst signal to vortex physics in the way one expects, cf. \cite{old}. 

A further interesting observable is the Lorenz ratio $L$. Previously in (\ref{eq:Lor1}) we found that $L \ll 1$ for the obvious reason that the electrical conductivity was large due to the long-lived supercurrent whereas the thermal conductivity was not. This simple argument will not apply with broken parity. We see in (\ref{eq:kap}) and (\ref{eq:ll}) that, due to the entropy carried by the vortices, the thermal conductivity is also enhanced by the fluctuating superconductivity. However, for the Lorenz ratio one needs the open circuit thermal conductivities. The matrix of open circuit conductivities is given by $\hat \kappa = \hat {\overline \kappa} - T \hat \alpha_+ \, \hat \rho \, \hat \alpha_-$.
As previously, hats denote matrices of conductivities. Using the formulae (\ref{eq:ss}) to (\ref{eq:ll}) for the conductivities one quickly finds that the fluctuating superfluid contributions precisely cancel, so that $\hat \kappa$ is not in fact enhanced. It follows that, with long-lived supercurrents,
\be
L \equiv \frac{\kappa}{\sigma T} \ll 1 \,, \qquad L^H \equiv \frac{\kappa^H}{\sigma^H T} \ll 1 \,.
\ee
This cancellation parallels that noted in \cite{Mahajan:2013cja} for the case of a long-lived momentum. The open circuit boundary conditions project out the long-lived modes from the thermal current.

The physics underlying the above formulae for the conductivities is straightforward. The first effect one sees is that -- in the absence of superfluid degradation, i.e. $\Omega = \Omega^H = 0$ --  the vortex quantities $s_v$ and $\rho_v$ have resulted in a proliferation of divergent transport coefficients. This is because external temperature gradients and electric fields now give rise to vortex current flow, according to (\ref{eq:jvcons2}), and the vortex current flow in turn couples to the non-dissipative supercurrent flow, according to (\ref{eq:jmod}). This effect and several others are transparent if the above results are recast in the language of the memory matrix. 

\subsection{Memory matrix description}

The formulae of the previous subsection can be recovered using the memory matrix approach. This works similarly to the discussion in section \ref{sec:memA} above, and once again leads to explicit microscopic formulae for the decay rates (now $\Omega$ and $\Omega^H$) as well as the incoherent conductivities. 

In the parity and time-reversal non-invariant case, it is instructive to consider the entire matrix of conductivities. The memory matrix formalism gives the conductivity $\sigma_{AB}$, where $A,B$ can be the $x$ or $y$ components of the total electrical or heat currents $\{J,J^Q\}$, as \cite{forster}
\be\label{eq:fullmemory2}
\sigma_{AB}(\omega) = \sum_{CD} \chi_{A C} \left(\frac{1}{- i \omega \chi +  M(\omega) + N}\right)_{CD} \chi_{D B} \,.
\ee
As previously, the sum runs over both the long lived operators and the hydrodynamic currents. The long lived operators are now $\{J_\phi^x, J_\phi^y\}$. That is, $\{C,D\} \in \{J,J^Q, J_\phi\}$. Both the $x$ and $y$ components of these operators appear.

The static susceptibilities are again given by equation (\ref{eq:chi}) and (\ref{eq:chi_memory}) above. Specifically, let\footnote{We have not allowed a $\delta^{ij}$ term in the final quantity $\chi_{J_\phi^i J_Q^j}$ in (\ref{eq:sus}). That is, the supercurrent does not directly `drag' a parallel thermal current. Such a coupling would violate the Josephson relation at zeroth order in the derivative expansion (leading to additional terms in (\ref{eq:jvcons2})), and is disallowed by gauge invariance.}
\be
 \chi_{J_\phi^i J_\phi^j} =  \frac{1}{\rho_s} \delta^{ij} \,, \qquad  \chi_{J_\phi^i J^j} = \frac{1}{m} \delta^{ij}  \,, \qquad  \chi_{J_\phi^i J_Q^j} = 0   \,. \label{eq:sus}
\ee
Even when parity is broken, terms proportional to $\epsilon_{ij}$ are forbidden by $\sf PT$.
These quantities are defined in the full theory with superfluid relaxation. 

The full formula (\ref{eq:fullmemory2}) is a little complicated, as it includes both the fluctuating superfluid modes and the incoherent contributions. As in section \ref{sec:memA} above, the formula becomes useful once we zoom into the physics of the slowly decaying superfluid excitations. We can imagine two different small parameters, $\vep$ and $\eta$, so that
\be\label{eq:NMscale}
M_{J_\phi J_\phi} \sim M_{J J_\phi} \sim \vep^2 \,, \qquad  N_{J_\phi J_\phi} \sim  N_{J J_\phi} \sim \eta \,.
\ee
We have already discussed the scaling of the components of $M$ around equation (\ref{eq:limit}) above and in the appendix. The new parameter $\eta$ quantifies the extent of time reversal symmetry breaking (without which, all the components of $N$ vanish). In order to bring out the physics in the cleanest possible way, we will take $\eta \sim \vep^2$. With this scaling, various effects arise at the same order in an $\vep \to 0$ expansion. In particular, taking $\omega \sim \vep^2$, combined with the above scalings, in (\ref{eq:fullmemory2}) leads precisely to the superfluid part of the hydrodynamic formulae for the thermoelectric conductivities (\ref{eq:ss}) through (\ref{eq:ll}) obtained above, with $\rho_v,\, s_v\to 0$ (since these terms are suppressed in the limit of weak parity breaking). One can also reproduce the incoherent hydrodynamic contributions at the same order in the scaling limit if the incoherent susceptibility is to taken to be large as in (\ref{eq:limit}) above. Furthermore, the inverse lifetime and oscillation frequency of the collective mode are now given by
\bea
\Omega & = & \rho_s M_{J_\phi^x J_\phi^x} \,, \\
\Omega^H & = & - \rho_s \left( M_{J_\phi^x J_\phi^y} + N_{J_\phi^x J_\phi^y} \right) \,.
\eea
In the low frequency scaling limit we have taken, $\omega$ may be set to zero in the memory matrix components $M_{J_\phi^i J_\phi^j}(\omega)$ appearing in the above formula.
The expression for $\Omega$ is essentially that obtained previously in (\ref{eq:key}) above, while the expression for $\Omega^H$ is new. Given an explicit mechanism for superfluid relaxation, these microscopic formulae can in principle be evaluated to obtain, for instance, the dc conductivities via (\ref{eq:dc_cond_PT}).

Beyond the dc resistivities, an interesting generic consequence of time-reversal-breaking fluctuating superconductivity -- seen for instance in (\ref{eq:ss}) and (\ref{eq:shsh}) -- is the existence of what we might call a `hydrodynamic supercyclotron' mode with complex frequencies
\be
\omega_\star = \pm \Omega^H - i \Omega \,.
\ee
Depending on the relative values of $\Omega$ and $\Omega^H$, however, a peak in the conductivity may or may not be visible at $\omega \sim \Omega^H$. Specifically, if $\Omega^H \leq \sqrt{3} \Omega$, the only visible feature is a Lorentzian-like peak at $\omega = 0$. This is illustrated in the figure \ref{fig:sw1} below.
\begin{figure}[h]
\centering
\includegraphics[height = 0.3\textheight]{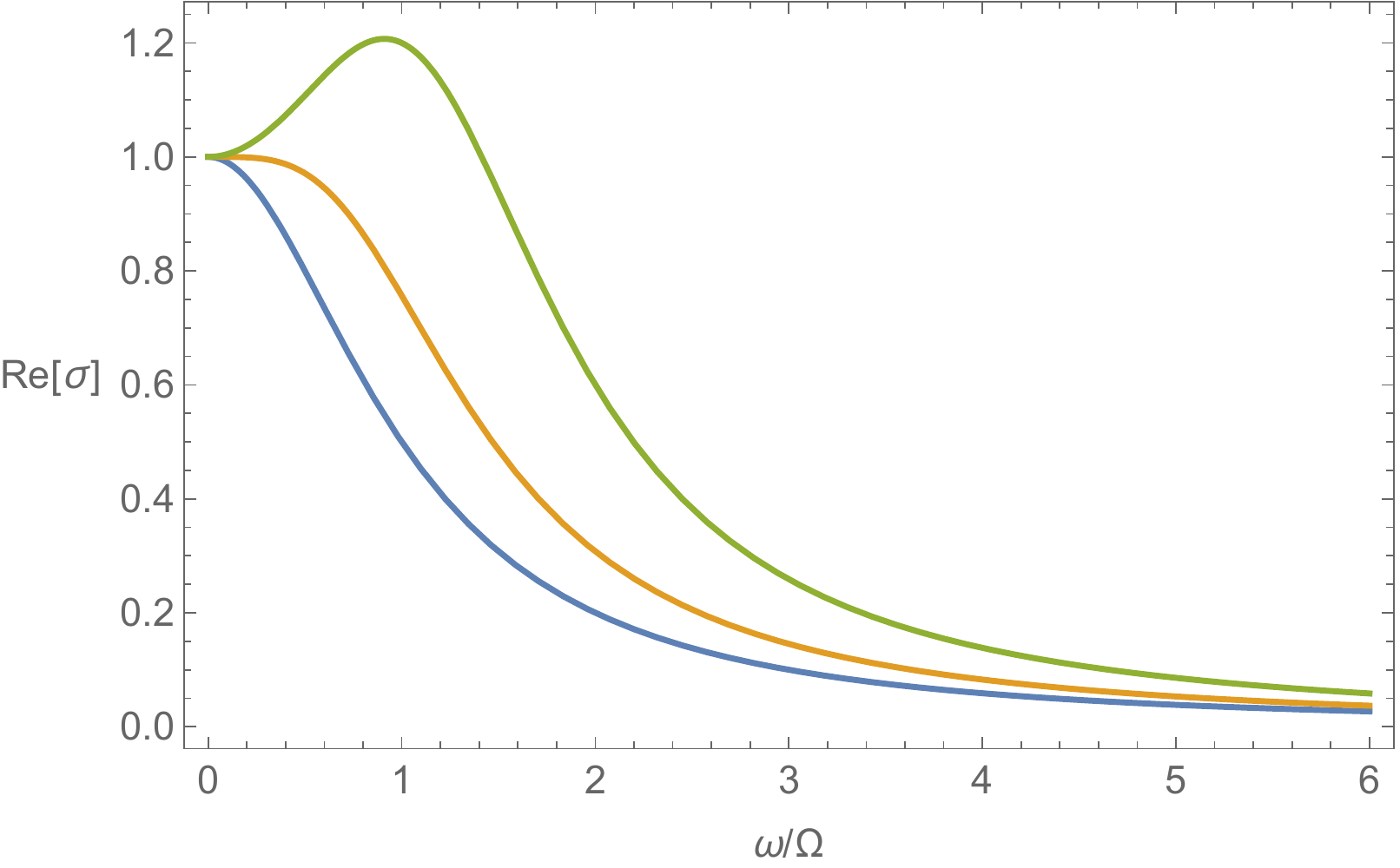}
\caption{\label{fig:sw1}  {\bf Optical conductivity for different values of $\Omega^H$}. From bottom to top: $\Omega^H = \{0,\Omega/\sqrt{3},\Omega\}$. In each plot $\Omega$ has been chosen so that $\sigma_\text{dc} = 1$, in units with $\rho_s /m^2 = 1$ in (\ref{eq:ss}), and we have set $\rho_v\to 0$}.
\end{figure}
Interestingly, the InO$_x$ optical conductivity data \cite{ac}, discussed in the introduction, potentially shows a flattening in the Drude-like peak and the possible emergence of a maximum away from zero frequency as the magnetic field is increased.

The following section considers a particular instance of a simple parity-violating interaction that degrades supercurrent.
This interaction will lead to $\eta \sim \vep$ -- rather than the $\eta \sim \vep^2$ assumed for illustrative purposes below (\ref{eq:NMscale}) -- and thus the hydrodynamics is a little different to that developed above. That is to say, we will have $M \ll N$. Furthermore, special features of this interaction will allow us to obtain more exact results for the conductivity than are possible in general.

\section{Supercurrent relaxation by a Chern-Simons interaction}
\label{sec:CS}

Consider the nonlocal and parity-violating density-current interaction
\be\label{eq:cs}
\Delta H = \frac{\lambda'}{2} \int \frac{d^2k}{(2\pi)^2} \frac{\rho_{-k} \left(\nabla \times j \right)_k^z}{k^2} + \text{h.c.} \,.
\ee
The $z$ superscript tells us to take the component orthogonal to the two dimensional plane. Here $\lambda'$ is a dimensionless coupling.

As before, we consider this interaction as a perturbation of the low energy effective description of the system in terms of superfluid hydrodynamics. There is a simple way that the new interaction (\ref{eq:cs}) can arise when parity is broken. Suppose that the low energy description contains an emergent $U(1)$ gauge field that couples to charged fields and furthermore has a Chern-Simons term. That is, we have the full Lagrangian
\be\label{eq:LagA}
\Lag = \Lag_\text{matter} + j_\mu (A^\mu + a^\mu) - \frac{1}{2 \lambda'} \epsilon^{\mu\nu\rho} a_\mu \pa_\nu a_\rho \,.
\ee
Here $A$ is the background electromagnetic field and $a$ the emergent gauge field. In this theory the electromagnetic current $j$ still corresponds to a global symmetry, while a linear combination of $j$ with the topological current $j_\text{top}^\mu \equiv \epsilon^{\mu\nu\rho} \pa_\nu a_\rho$ is gauged. Integrating out the emergent gauge field (this requires gauge fixing to invert the propagator, as usual) generates the `Hopf interaction' \cite{Zee:1996fe}
\be\label{eq:local}
\Lag =  \frac{\lambda'}{2} j_\mu \frac{\epsilon^{\mu
\nu\rho} \pa_\rho}{\pa^\sigma \pa_\sigma} j_\nu \,.
\ee
The non-relativistic limit of this interaction gives the Hamiltonian (\ref{eq:cs}), together with a current-current interaction that will not relax the superfluid efficiently.

The `Chern-Simons interaction' (\ref{eq:cs}) causes a time dependence in the supercurrent operator according to\footnote{The result (\ref{eq:com2}) comes from the commutator of the supercurrent with the density operator $\rho$ in the Hamiltonian (\ref{eq:cs}). The commutator of the supercurrent with the current operator $j$ in (\ref{eq:cs}) -- where $j$ can be taken to have the form $j = m^{-1} \left(1 + \alpha \, \rho + \beta \, \rho^2 +  \cdots \right) \nabla \phi$ -- is subleading. This is because the terms in $j$ that we have just written that involve $\rho$ are nonlinear in hydrodynamic variables. Hydrodynamic correlators obey Gaussian factorization, and thus, upon taking the correlator (\ref{eq:key}) of $[\Delta H, J_\phi]$ to obtain the decay rate, nonlinear effects are suppressed by factors of (small) momenta times the correlation length. That is to say, we can neglect nonlinear terms for the same reason that we can focus on linearized hydrodynamics to obtain Green's functions.}
\be\label{eq:com2}
i[\Delta H, J^i_\phi] = - \frac{\lambda'}{m}  \lim_{k \to 0} \epsilon^{ij} j^{T\,j}\,.
\ee 
Here $j^T$ is the transverse part of the electrical current, satisfying $\nabla \cdot j^T = 0$.
While used in deriving (\ref{eq:com2}), the distinction between longitudinal and transverse is not important at the end of the day because the $k=0$ mode of the current can be considered as either longitudinal or transverse (strictly, it is the harmonic part of the current).

As in the previous case of equation (\ref{eq:H1J}), technically the right hand side of (\ref{eq:com2}) should be the integral of the current outside of vortices. However, the essential difference with (\ref{eq:H1J}) is that $\dot J_\phi$ is not a total derivative in this case. The relaxation rate will be dominated by the contribution from throughout the superfluid rather than vortex cores. This will allow, below, the relaxation rate to be evaluated in terms of universal quantities that appear in the superfluid hydrodynamics.

The result (\ref{eq:com2}) can be understood physically from the Chern-Simons term for the emergent gauge field (\ref{eq:LagA}). In particular, this perspective clarifies the role of vortices. This Chern-Simons term has two effects. The equations of motion following from (\ref{eq:LagA}) tie together the charge density with an emergent magnetic field $b$ and the current density with an emergent electric field $e$:
\be\label{eq:BE}
b(x) = \lambda' \, \rho(x) \,, \qquad e^i(x) = - \lambda' \, \epsilon^{ij} j^j(x) \,.
\ee
The second of these equations is essentially (\ref{eq:com2}): an electric current creates a transverse emergent electric field, which is in turn equivalent to a time-dependent phase gradient (as the emergent gauge potential will also now appear in the Josephson relation). As we noted above, topologically speaking, we expect a time-dependent phase gradient should involve vortex flow. The first equation in (\ref{eq:BE}) shows this explicitly via the following steps. Firstly, electric current is of course the flow of charge density (via the conservation law $\dot \rho + \nabla \cdot j = 0$). The first equation in (\ref{eq:BE}) shows that a flow of charge necessitates a flow of emergent magnetic flux. But this flux can only penetrate the superconductor by creating a vortex. Therefore, current flow is accompanied by vortex flow. The beautiful fact about the first equation in (\ref{eq:BE}) is that it ties the presence of vortices directly to the hydrodynamic variable $\rho$. Thus, unlike in the case discussed in section \ref{sec:BKT}, which required additional input from e.g. BKT theory to obtain the free vortex density $n_F$, the computation of $\Omega$ and $\Omega^H$ in the following sections will be self-contained within hydrodynamics.

Something close to the dynamics described above is realized in recent
theories of the metallic `vortex liquid' state \cite{mikeandsri,gal}. It is also interesting to note that the theory of anyon superconductivity (e.g. \cite{anyon}) contains a non-relativistic Chern-Simons term that imposes the first but not the second of the equations in (\ref{eq:BE}). Therefore, supercurrent is not relaxed in that theory (at least, not by this mechanism). Superfluid hydrodynamics with a topological term analogous to (\ref{eq:local}), but higher order in derivatives, was considered in \cite{Golkar:2014paa}. That term does not relax the total supercurrent.

We will obtain the conductivity by two distinct methods. Firstly, from the memory matrix formalism together with the interaction (\ref{eq:cs}), that leads to superfluid relaxation via (\ref{eq:com2}). Second, by incorporating the effects of the emergent Chern-Simons gauge field directly into hydrodynamics. 

\subsection{Conductivities from the memory matrix}
\label{sec:memD}

The memory matrix expression for the conductivity is again (\ref{eq:fullmemory2}). Several simplifications and special features occur in the case of Chern-Simons relaxation. This is because the time derivative of the supercurrent in (\ref{eq:com2}) is itself a hydrodynamic variable (the total electrical current). We can then do the following. Firstly, restrict attention to the electric conductivities, so that the external $\{A,B\}$ indices in (\ref{eq:fullmemory2}) only run over the $x$ and $y$ components of the electric current $J$. The summed-over indices $\{C,D\}$ now need only run over the $x$ and $y$ components of $\{J_\phi,J\}$. It follows from the expression for the memory matrix in (\ref{eq:MMM}) that all components of the memory matrix involving $J_\phi$ vanish. This is because, from (\ref{eq:com2}), $\dot J_\phi \sim J$, but the projector ${\mathcal Q}$ projects out both components of the $J$ operator, and hence ${\mathcal Q} | \dot J_\phi ) = 0 = ( \dot J_\phi | {\mathcal Q}$. That is, the memory matrix takes the form
\be \label{eq:memmat_CS}
M_{J^i J^j} \neq 0 \,, \qquad M_{J^i_\phi J^j} = M_{J^i J_\phi^j}  = M_{J_\phi^i J_\phi^j}  = 0 \,.  
\ee
Note the difference with the more typical case considered in section \ref{sec:noP}. The components of the memory matrix that typically control slow relaxation are in fact zero in this case.

The matrix $N$ is computed in this case directly from the definition (\ref{eq:NN}) and the formula (\ref{eq:com2}) for $\dot J_\phi$. We have (using isotropy and asymmetry of $N$)
\be
N_{J^i J^j} = N_{J^x J^y} \, \epsilon^{ij} \,, \qquad N_{J^i_\phi J^j_\phi} = \frac{\lambda'}{m^2} \epsilon^{ij} \,, \qquad
N_{J^i J_\phi^j} = - N_{J_\phi^j J^i} = - \frac{\lambda'}{m} \epsilon^{jk} \chi_{J^i J^k} \,.
\ee

The matrix of susceptibilities takes the form
\be
\chi_{J^i J^j} = \chi_{J^x J^x} \, \delta^{ij} \,, \qquad 
\chi_{J^i J_\phi^j} = \chi_{J_\phi^j J^i} = \frac{1}{m} \delta^{ij} \,, \qquad 
\chi_{J_\phi^i J_\phi^j} = \frac{1}{\rho_s} \delta^{ij} \,. \label{eq:matt}
\ee
To set the $xy$ components to zero we firstly used isotropy and secondly, for $\chi_{J^x J_\phi^y}$, noted that
$0 = N_{J^x J^x} = \frac{- \lambda'}{m} \chi_{J^x J_\phi^y} \,.$
All of the above quantities, $M$, $N$ and $\chi$, are exact in $\lambda'$ at this point, as all nonzero quantities that appear, $\{M_{J^i J^j}, N_{J^x J^y}, \chi_{J^x J^x}, m, \rho_s\}$, are evaluated in the theory with $\lambda'$. 

Inserting the above expression in the memory matrix formula (\ref{eq:fullmemory2}) and taking the d.c.~ limit $\omega \to 0$ one obtains
\be\label{eq:halleasy}
\sigma = 0 \,, \qquad \sigma^H = - \frac{1}{\lambda'} \,.
\ee
This result is exact in $\lambda'$, i.e. we do not need to take $\lambda'$ small. In fact, the result (\ref{eq:halleasy}) has nothing to do with fluctuating superfluidity. It follows directly from the Chern-Simons Lagrangian (\ref{eq:LagA}), with no assumptions about whether the system is superfluid or not. To see this we can shift the emergent gauge field in (\ref{eq:LagA}) by $a \to a - A$. So it is a good thing that the memory matrix reproduces this result. Fluctuating superfluidity, however, leaves a strong imprint on the frequency-dependent conductivity, as we now see.

The inverse matrix in (\ref{eq:fullmemory2}) leads to a complicated $\omega$ dependence. The interesting physics we wish to zoom in on is the resonance that appears at small $\omega$ when $\lambda'$ is small. Therefore we take the following scaling limit of (\ref{eq:fullmemory2}) with as $\lambda' \to 0$:
\be\label{eq:scallim}
\omega \sim \lambda' \,, \qquad \chi_{\rm inc} \equiv \chi_{J_xJ_x} - \frac{\rho_s}{m^2} \sim \frac{1}{\sqrt{\lambda'}} \,.
\ee
As in the previous discussion around (\ref{eq:limit}) above, the second of these two scalings is not essential. However, it makes various expressions physically more transparent, allowing the `incoherent' contribution to appear at the same order as the fluctuating superconductivity. In the present context, it also results in the width and location of the collective `super-cyclotron' mode scaling in the same way with $\lambda'$. Otherwise, the frequency (energy gap) of the mode is much greater than its inverse lifetime.

In this scaling limit (\ref{eq:scallim}) we obtain from (\ref{eq:fullmemory2})
\begin{align}
\sigma
	&= - \frac{m^2}{\lambda'^{\,2} \rho_s}\frac{\omega(\omega\Omega+i(\Omega^2 + \Omega_H^2))}{(-i\omega+\Omega)^2 +\Omega_H^2} + \mathcal{O}\left((\lambda')^0\right)\, , \label{eq:a1} \\
\sigma^H
	&= - \frac{1}{\lambda'}- \frac{m^2}{\lambda'^{\,2}\rho_s}\frac{\omega^2\, \Omega_H}{(-i\omega+\Omega)^2 +\Omega_H^2} + \mathcal{O}\left((\lambda')^0\right) \, ,\label{eq:a2} 
\end{align}
with
\begin{align}
\Omega
	&= \frac{\lambda'^{\,2}\rho_s}{m^2} \frac{\chi_{\rm inc}^2M_{J^xJ^x}}{M_{J^xJ^x}^2 + (M_{J^xJ^y} + N_{J^xJ^y})^2} \\
	&\quad \equiv \frac{\lambda'^{\,2}\rho_s}{m^2} \frac{\sigma_0}{(1-\lambda' \sigma_0^H)^2 + (\lambda' \sigma_0)^2}\, , \label{eq:a3} \\
\Omega_H
	&=\frac{\lambda'\rho_s}{m^2}  \left(1-\lambda'\frac{\chi_{\rm inc}^2 (M_{J^xJ^y} + N_{J^x J^y})}{M_{J^xJ^x}^2 + (M_{J^xJ^y} + N_{J^xJ^y})^2} \right) \\
	& \quad \equiv\frac{\lambda'\rho_s}{m^2}\frac{1-\lambda' \sigma_0^H}{(1-\lambda' \sigma_0^H)^2 + (\lambda' \sigma_0)^2}\, , \label{eq:a4} 
\end{align}
where to get the second equalities we identified the matrix $(\sigma_0)^{ij} = \sigma_0 \delta^{ij} + \sigma_0^H \epsilon^{ij}$ of incoherent conductivities as the inverse of the incoherent resistivity matrix
\be
(\rho_0)_{ij} \equiv \frac{M_{J^i J^j} + N_{J^i J^j}}{\chi_{\rm inc}^2}  - \lambda' \epsilon^{ij}\,.
\ee
Note that the $M_{J^i J^j}$ appearing in the above expressions are all evaluated at $\omega = 0$. Any higher order in $\omega$ corrections are subleading in the limit (\ref{eq:scallim}).

The most distinctive feature of the above expressions is the appearance of what we have called a `hydrodynamic supercyclotron' mode at frequencies
\be
\omega_\star = \pm \Omega^H - i \Omega = \frac{\lambda' \rho_s}{m^2} \frac{\pm 1 - \lambda' (\pm \sigma_0^H + i \sigma_0)}{(1-\lambda' \sigma_0^H)^2 + (\lambda' \sigma_0)^2} = \frac{\lambda' \rho_s}{m^2} \frac{1}{\pm 1 - \lambda' (\pm \sigma_0^H - i \sigma_0)} \,.\label{eq:scc}
\ee
This mode has some similarities with the hydrodynamic cyclotron resonance discovered in \cite{Hartnoll:2007ih, Hartnoll:2007ip} and further investigated in \cite{Lucas:2015pxa}. In particular, in both cases, the lifetime of the mode depends upon the `incoherent' conductivity $\sigma_0$. However, the underlying physics is quite different. The supercyclotron mode above arises due to the motion of a superfluid condensate that has become phase-disordered due to the dynamics of vortices that carry magnetic flux of the emergent Chern-Simons field. To emphasize the formal analogy, however, in Appendix \ref{sec:mag} we rederive the general magnetohydrodynamic results of \cite{Lucas:2015pxa} using the same memory matrix manipulations as have been performed in this section.

As we found previously, the supercyclotron mode is only visible as a feature in the optical conductivity if $\Omega^H > \Omega/\sqrt{3}$. The optical conductivity for Chern-Simons relaxation is shown in figure \ref{fig:sw2} below.
\begin{figure}[h]
\centering
\includegraphics[height = 0.28\textheight]{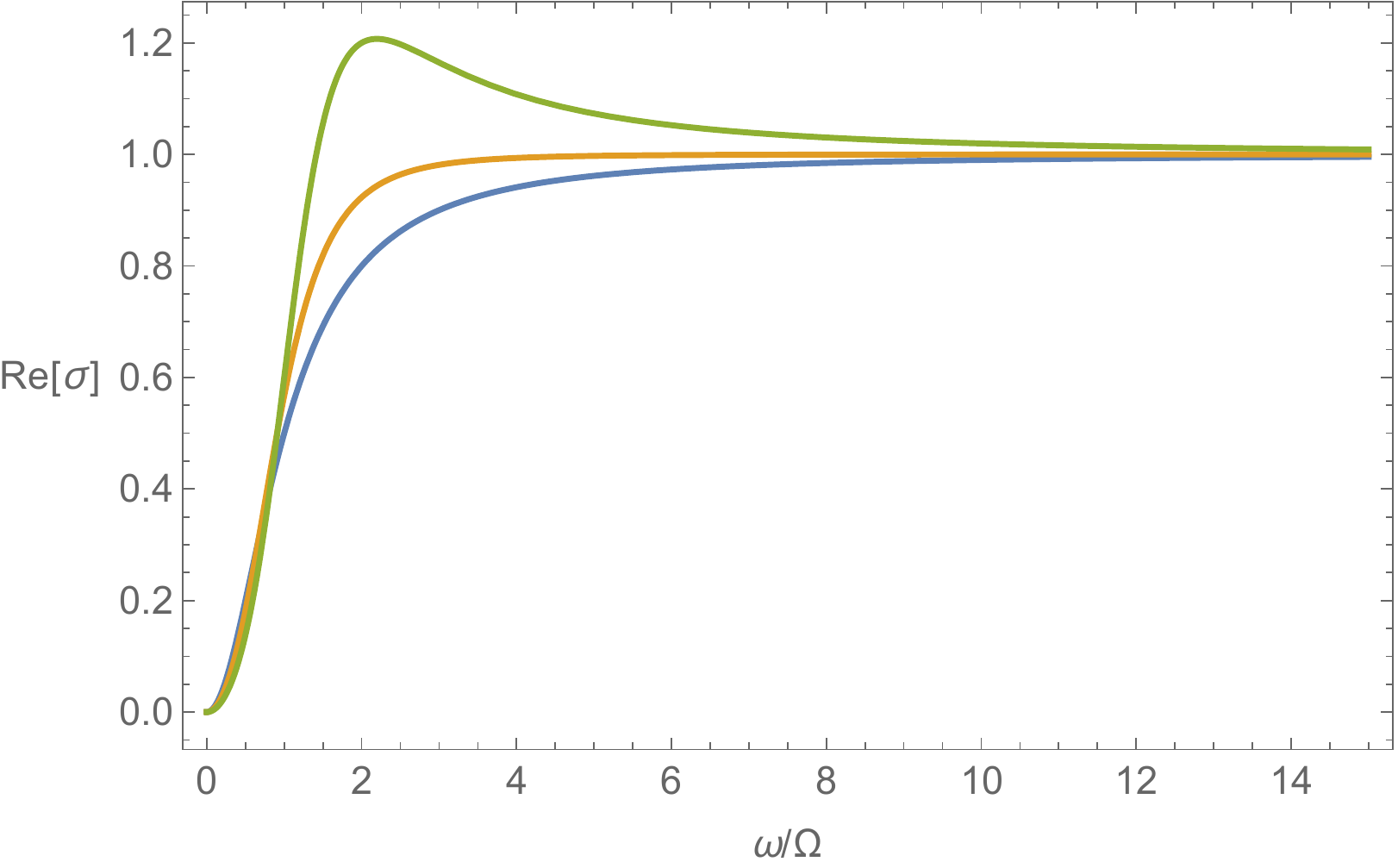}
\caption{\label{fig:sw2}  {\bf Chern-Simons optical conductivity for different values of $\Omega^H$}. From bottom to top: $\Omega^H = \{0,\Omega/\sqrt{3},\Omega\}$. In each plot $\Omega$ has been chosen so that $\sigma(\omega\to\infty) = 1$, in units with $m^2/(\lambda'^2\rho_s) = 1$ in (\ref{eq:a1}).}
\end{figure}

Due to the particular features of the Chern-Simons interaction, in this case the collective phase-fluctuation mode (\ref{eq:scc}) does not determine the dc conductivities (\ref{eq:halleasy}).  In particular, the physics described in this section seems not to be the dominant phase-relaxing dynamics visible in the InO$_x$ data of \cite{ac}, that finds nonvanishing longitudinal dc conductivities, as well as $\Omega \gg \Omega^H$. Nonetheless, the phase of mater we have just characterized, a ``topologically ordered superfluid vortex liquid'', seems to involve plausible ingredients and will hopefully arise in other contexts. Finally, generalizations of the interaction (\ref{eq:cs}) are likely to exist, involving for instance the energy rather than the charge current.

\subsection{Chern-Simons superfluid hydrodynamics}

An alternative method to obtain the conductivities is to study the hydrodynamics of a superfluid coupled to an emergent $U(1)$ gauge field with a Chern-Simons term, as described by (\ref{eq:LagA}). Instead of integrating out the emergent gauge field $a$ from the start to produce a Hopf term, as was done in the previous section, $a$ is incorporated in the hydrodynamics with the replacement $A\to A^{\rm tot}=A+a$. The effect of the Chern-Simons interaction is then accounted for by putting $a_\mu$ on shell in the constitutive relations and Josephson equation:
\begin{equation}
\epsilon^{\mu\nu\rho}\partial_\nu a_\rho = \lambda' j^\mu\, .
\end{equation}
Concretely, this amounts to revisiting the hydrodynamics (\ref{eq:jjj}), (\ref{eq:qqq}) with the following replacement
\begin{equation}
E_i = -\nabla_i \mu \quad\longrightarrow \quad
E^{\rm tot}_i = E_i + e_i = -\nabla_i\mu -\lambda' \epsilon_{ij} j_j\, .
\end{equation}
In this way, the constitutive relations and the Josephson relation become
\begin{eqnarray}
j^i -  \frac{\bar\rho_s}{m} \left(\delta^{ij} - \rho_v \epsilon^{ij} \right) u^j_\phi  & =  &- \hat \sigma^{ij}_0 (\nabla^j \mu+\lambda' \epsilon^{jk} j^k) - \hat \alpha^{ij}_0 \nabla^j T  + \cdots  \,, \label{eq:jjjCS} \\
{\textstyle \frac{1}{T}}j^{Q\,i} + \frac{\bar\rho_s}{m} s_v \epsilon^{ij} u^j_\phi  & = & - \hat \alpha^{ij}_0 (\nabla^j \mu+\lambda' \epsilon^{jk} j^k) - (\hat \kappa_0^{ij}/T) \nabla^j T + \cdots \,, \label{eq:qqqCS} \\
j^i_v  & = & - \frac{s_v}{2\pi} \nabla^i T - \frac{\rho_v}{2\pi} \left(\nabla^i  \mu +\lambda' \epsilon^{ik} j^k \right) - \gamma \nabla^i n_v + \cdots \,, \\
m \pa_t u^i_\phi & = & - \nabla^i \mu+ \epsilon^{ij} (2\pi j^j_v - \lambda' j^j) + \cdots \,.
\end{eqnarray}
We have chosen to express the constitutive relations directly in terms of the matrices of incoherent conductivities $\hat \sigma_0, \hat \alpha_0, \hat \kappa_0$ rather than the diffusivities that were used in (\ref{eq:jjj}), (\ref{eq:qqq}). These are related via the Einstein relations (\ref{eq:incoh}). A bar has been placed over $\bar \rho_s$ as it will turn out shortly that, due to the extra $j_i$ terms appearing in various places on the right hand side of the above equations, this quantity is no longer the superfluid density as defined via the susceptibility (\ref{eq:matt}).

The electrical conductivities can then be obtained from the hydrodynamic equations of motion in the same way that we have done several times in this paper. The answers are
\begin{align}
\sigma_{xx} 
	&= -\frac{m^2}{\lambda'^2 \bar \rho_s (1+\rho_v^2)}\frac{\omega(\omega \Omega + i(\Omega^2 + \Omega_H^2))}{(-i\omega + \Omega)^2 + \Omega_H^2}\, , \\
\sigma_{xy} 
	&= -\frac{1}{\lambda'} - \frac{m^2}{\lambda'^2 \bar \rho_s (1+\rho_v^2)}\frac{\omega^2\Omega_H}{(-i\omega + \Omega)^2 + \Omega_H^2}\, ,
\end{align}
with
\begin{align}
\Omega 
	&= \frac{\lambda'^2 \bar \rho_s(1+\rho_v^2)}{m^2}\frac{\sigma_0}{(1-\lambda' \sigma_0^H)^2 + (\lambda' \sigma_0)^2}\, , \label{eq:last}\\
\Omega_H 
	&= \frac{\lambda' \bar \rho_s(1+\rho_v^2)}{m^2}\frac{1-\lambda'\sigma_0^H}{(1-\lambda' \sigma_0^H)^2 + (\lambda'\sigma_0)^2}\, .
\end{align}
These agree precisely with the memory matrix answers of the previous subsection -- equations (\ref{eq:a1}), (\ref{eq:a2}), (\ref{eq:a3}), (\ref{eq:a4}) --  upon making the identification: $\rho_s = \bar \rho_s (1+\rho_v^2)$.

\section{Discussion}
\label{sec:discuss}

We will end with a few brief comments on the results we have obtained.

Much of our discussion has been phrased in terms of hydrodynamics. By hydrodynamics we mean the long wavelength dynamics of conserved quantities and Goldstone bosons. However, the driving motor behind our main results is the memory matrix formalism. We could have dispensed with hydrodynamics altogether. We have kept the hydrodynamic perspective because it may be more familiar to readers and is arguably physically more transparent. We have seen, however, that in order to cleanly reproduce hydrodynamics, including the incoherent contributions, one needs to take a certain scaling limit of the memory matrix expressions. The memory matrix expressions are exact in the first instance, and one can then see unambiguously what approximations are required to recover the hydrodynamic answers. The memory matrix is a systematic tool for capturing the effects of long-lived excitations in a system.

To obtain the behavior of observables for specific phase-fluctuating systems -- such as the temperature and magnetic field dependence of the dc conductivities -- one needs firstly to know the supercurrent-relaxing interaction $\Delta H$. We have investigated two such interactions, but other possibilities exist. In the cases we considered, the charge interacted with itself (as in the short range interaction we considered) or the electrical current (as in the Chern-Simons interaction we considered). However, one can imagine interactions of the charge density with other operators, of the form $\Delta H \sim \int d^2x \, \rho(x) \ocal(x)$, for some local operator $\ocal(x)$. This operator will then take the place of the charge density in the formula (\ref{eq:Omega1}) for the supercurrent relaxation rate. Once the interaction itself is given, one must furthermore determine, for instance, the temperature dependence of the thermodynamic susceptibilities and other quantities appearing in formulae such as (\ref{eq:OO}) or (\ref{eq:last}). However, even when this temperature dependence is not known, the formalism we have developed ties together a collection of distinct observables in terms of just a few quantities.

Among our more generic results is the prediction that the optical conductivity of phase-disordered superconductors with broken parity should reveal a `hydrodynamic supercyclotron' mode at complex frequencies $\omega_\star = \pm \Omega^H - i \Omega$. We noted that if $\Omega^H$ is small (relative to $\Omega$) the peak in the optical conductivity will still be centered at zero frequency, but will exhibit deviations from a simple Lorentzian form. This mode should be accessible to standard experimental probes.

Finally, an important scenario we have not considered is one in which the superconducting state is close to a percolation phase transition, with large normal state domains across the sample. In this case, quantum tunneling of vortices between closely spaced normal state domains gives a mechanism for supercurrent relaxation. This is distinct from the setup of section \ref{sec:BKT}, as it does not require a density of free vortices. The vortices now only appear as virtual tunneling events. For this reason, it is a promising framework for relaxing the supercurrent even at zero temperature. Transport through such `quantum melts' has been discussed in e.g. \cite{assa}. It will be interesting to re-investigate this scenario from the perspective of the memory matrix that we have developed. Such a calculation will be similar to the one-dimensional memory matrix computation of \cite{andrei}, as the tunneling occurs in the thin necks of superconductivity separating normal domains.

\section*{Acknowledgements}

It is a pleasure to acknowledge helpful discussions with Ehud Altman, Peter Armitage, Aharon Kapitulnik, Steve Kivelson, Andrew Lucas, Akash Maharaj, Tyler Merz, Djordje Radicevic and Leo Radzihovsky. The work of B.G. is supported by the Marie Curie International Outgoing Fellowship nr 624054 within the 7th European Community Framework Programme FP7/2007-2013. The research of S.A.H. is partially supported by an Early Career Award from the US Department of Energy. The work of R.D. is supported by the Gordon and Betty Moore Foundation EPiQS Initiative through Grant GBMF\#4306.
 
\appendix


\section{Proofs of memory matrix statements}
\label{sec:Qe}

\subsection{Time reversal invariant case}

In the presence of time-reversal symmetry, and for operators with the same sign under time-reversal, the memory matrix result (\ref{eq:fullmemory}) can be somewhat simplified. First, $N$ vanishes since it measures an overlap of two operators of opposite sign under time-reversal. Second, the Liouville operator $L=[H,{\bigcdot} \, ]$ must act an even number of times in $M(z)$ (\ref{eq:MMM}), so that
\begin{equation}\label{app_M_with_T}
M_{CD}(\omega) = \frac{i}{T\omega} \left(\dot C \left|\frac{1}{1-L \mathcal Q L/\omega^2}\right| \dot D\right)\, ,
\end{equation}
where we also noted that the projection operator
\begin{equation}\label{app_Q}
\mathcal Q= 1 - \frac{1}{T}\sum_{A,B} |A)\chi^{-1}_{AB}(B|
\end{equation}
can be set to 1 when acting on $\dot C$ or $\dot D$, by time-reversal symmetry. Since the hermitian operator $L$ is anti-symmetric and $\mathcal Q$ is symmetric, (\ref{app_M_with_T}) implies that here $M_{CD}(\omega)$ is symmetric.

Now in the case of interest we consider the operators $\{J,J_\phi\}$, with
\begin{equation}\label{app_Jphidot}
\dot J_\phi = i L J_\phi = i(L_0 + \varepsilon L_1)J_\phi = i\varepsilon L_1 J_\phi\, ,
\end{equation}
whereas $\dot J$ is generically of order unity. This implies
\begin{equation}
M_{JJ} \sim 1 \, , \qquad M_{J_\phi J_\phi} \sim \varepsilon^2\, , \qquad M_{JJ_\phi} \sim \varepsilon^2\, ;
\end{equation}
the first two relations are direct consequences of (\ref{app_Jphidot}), and we now prove the third.
Defining 
\begin{equation}
L_{J_\phi} = [J_\phi,{\bigcdot} \, ]
\end{equation}
and noticing that 
\begin{equation}
L|J_\phi) = \left|[H,J_\phi]\right) = -L_{J_\phi}|H)\, ,
\end{equation}
one has
\begin{equation}
\begin{split}\label{eq:MJJphisuppressed}
M_{JJ_\phi}(\omega) 
	&= \frac{i}{T\omega}\left( J \left|L_0\frac{1}{1-L_0 \mathcal Q L_0/\omega^2}L\right|  J_\phi\right) + {\mathcal O}(\varepsilon^2) \\
	&= \frac{i}{T\omega}\left( J \left|L_0\frac{1}{1-L_0 \mathcal Q L_0/\omega^2}L_{J_\phi}\right| -\varepsilon H_1\right) + {\mathcal O}(\varepsilon^2) \\
	&= \frac{i}{T\omega}\left( J \left|L_{J_\phi}L_0\frac{1}{1-L_0 \mathcal Q L_0/\omega^2}\right| -\varepsilon H_1\right) + {\mathcal O}(\varepsilon^2) \\
	&= 0 + {\mathcal O}(\varepsilon^2)\, ,
\end{split}
\end{equation}
where we used the fact that $L_{J_\phi}$ commutes with both $L_0$ and $\mathcal Q$, and the last step follows because $J$ carries no winding of the superfluid phase, so\footnote{This can be seen explicitly: isotropy and parity require $[J^j,J_\phi^k]  = \frac{1}{2}[J^i,J_\phi^i] \delta^{jk}$ and using the Ward identity one has
\begin{equation*}
[J^i,J_\phi^i] 
	= \int d^2 x d^2 y \, \partial_{y^i}[j^i(x),\phi(y)]
	= -\int d^2 x d^2 y \, \partial_{x^i}[j^i(x),\phi(y)]
	= \int d^2 x d^2 y \, \partial_{t}[\rho(x),\phi(y)] = 0\, .
\end{equation*}
}
\begin{equation}
L_{J_\phi} J = [J_\phi,J] = 0\, .
\end{equation}

The memory matrix formula (\ref{eq:fullmemory}) can now be used to find the small imaginary pole in the conductivity
\begin{equation}\label{app_pole}
\Omega \simeq \rho_s\lim_{\omega\to 0} M_{J_\phi J_\phi}(\omega) \simeq \rho_s \lim_{\omega\to 0} \frac{{\rm Im\,} G_{\dot J_\phi \dot J_\phi}(\omega)}{\omega} \, .
\end{equation}
The last step was accomplished by setting the remaining projection operator in (\ref{app_M_with_T}) to $\mathcal Q\to 1$. Although this is not obvious from e.g.\ time-reversal symmetry alone, it is in fact correct to leading order in $\varepsilon$, as we now show. Let $\widetilde M_{J_\phi J_\phi}$ be the matrix element evaluated with $\mathcal Q\to 1$. First notice that the definition (\ref{app_Q})  of $\mathcal Q$  implies 
\begin{equation}
L\mathcal Q L = L^2 - \frac{|\dot J)(\dot J|}{\chi_{\rm inc}} + {\mathcal O}(\varepsilon)\, .
\end{equation}
Using the short-hand notation
\begin{equation}
X = 1 - L^2/\omega^2\, , \qquad
Y = \frac{|\dot J)(\dot J|}{T\omega^2\chi_{\rm inc}} \, ,
\end{equation}
one can write
\begin{equation}\label{app_XYtricks}
\begin{split}
M_{JJ}(\widetilde M_{J_\phi J_\phi}-M_{J_\phi J_\phi})
	&= \frac{i}{T\omega} \left(\dot J \left|\frac{1}{X+Y}\right| \dot J\right)
	\frac{i}{T\omega} \left(\dot J_\phi \left|\frac{1}{X}-\frac{1}{X+Y}\right| \dot J_\phi\right)\\
	&= \frac{i}{T\omega} \left(\dot J \left|\frac{1}{X+Y}\right| \dot J\right)
	\frac{i}{T\omega} \left(\dot J_\phi \left|\frac{1}{X}Y\frac{1}{X+Y}\right| \dot J_\phi\right)\\
	&= \frac{i}{T\omega} \left(\dot J \left|\frac{1}{X+Y}\right| \dot J\right)
	\frac{i}{T\omega} \left(\dot J_\phi \left|\frac{1}{X}\frac{|\dot J)(\dot J|}{T\omega^2\chi_{\rm inc}}\frac{1}{X+Y}\right| \dot J_\phi\right)\\
	&= \frac{i}{T\omega} \left(\dot J_\phi \left|\frac{1}{X}\frac{|\dot J)(\dot J|}{T\omega^2\chi_{\rm inc}}\frac{1}{X+Y}\right| \dot J\right)
	\frac{i}{T\omega} \left(\dot J \left|\frac{1}{X+Y}\right| \dot J_\phi\right)\\
	&= \frac{i}{T\omega} \left(\dot J_\phi \left|\frac{1}{X}Y\frac{1}{X+Y}\right| \dot J\right)
	\frac{i}{T\omega} \left(\dot J \left|\frac{1}{X+Y}\right| \dot J_\phi\right)\\
	&= \frac{i}{T\omega} \left(\dot J_\phi \left|\frac{1}{X}-\frac{1}{X+Y}\right| \dot J\right)
	\frac{i}{T\omega} \left(\dot J \left|\frac{1}{X+Y}\right| \dot J_\phi\right)\\
	&= M_{JJ_\phi}(\widetilde M_{JJ_\phi} - M_{JJ_\phi})\, ,
\end{split}
\end{equation}
where the algebraic identity
\begin{equation}
\frac{1}{X}-\frac{1}{X+Y} = \frac{1}{X}Y\frac{1}{X+Y}
\end{equation}
was used twice. Since the right-hand side in (\ref{app_XYtricks}) is ${\mathcal O}(\varepsilon^3)$, we have
\begin{equation}
M_{J_\phi J_\phi} =  \widetilde   M_{J_\phi J_\phi} + {\mathcal O}(\varepsilon^3)\, ,
\end{equation}
showing explicitly that one can take $\mathcal Q\to 1$ in Eq.~(\ref{app_pole}), to leading order in $\varepsilon$.


\subsection{Non-time reversal invariant case}
 
Without time reversal symmetry the projection operator $\mathcal Q$ cannot be set to 1 (even perturbatively), as illustrated in Chern-Simons relaxation (\ref{eq:memmat_CS}) where $\mathcal Q$ entirely cancels $M_{JJ_\phi}$ and $M_{J_\phi J_\phi}$. However it is still true that $M_{JJ_\phi} \sim \varepsilon^2$, since the steps in Eq.~(\ref{eq:MJJphisuppressed}) can be carried out with the general form (\ref{eq:MMM}) for $M$ -- all that is needed is that the operator $L_{J_\phi}$ commutes with both $L_0$ (supercurrent conserved in the original theory) and with $\mathcal Q$, which follows from $[J_\phi,J]=0$.

 
\section{Magnetotransport revisited}
\label{sec:mag}

The memory matrix method used in the text can also be applied to magnetotransport \cite{Lucas:2015pxa}. Here we will rederive some results from \cite{Lucas:2015pxa} using manipulations very similar to those of section \ref{sec:memD} in the main text.

In magnetotranport, the role of the superfluid current is played by momentum, which is slowly relaxed by a small magnetic field according to:
\begin{equation}\label{eq:P_relax}
\dot P^i = B \epsilon^{ij} J^j + \ocal (B^2)\, .
\end{equation}
The order $B^2$ term arises if $P$ is the gauge invariant momentum \cite{Lucas:2015pxa}, and only gives subleading contributions to the expressions below.
This relaxation of momentum will resolve the delta function in the conductivities, as can be seen by using the memory matrix formula (\ref{eq:fullmemory2}) where the indices $\{C,D\}$ now run over the operators $\{J,P\}$. Since the projector $\mathcal Q$ projects out $J$, the components of the memory matrix take the form
\begin{equation}
M_{J^iJ^j}\neq 0\, , \qquad
M_{J^iP^j}= 0\, , \qquad
M_{P^iP^j}= 0\, .
\end{equation}
The susceptibilities are given by (these are the definitions of $Q$ and $\mathcal M$)
\begin{equation}
\chi_{J^i J^j}= \chi_{J^x J^x}\,\delta^{ij}\, , \qquad 
\chi_{J^i P^j}=  Q\, \delta^{ij}\, , \qquad 
\chi_{P^i P^j}= \mathcal M\, \delta^{ij}\, ,  
\end{equation}
where the relaxation equation (\ref{eq:P_relax}) forces $\chi_{J_x P_y}=0$ (the same way that we noted a certain susceptibility was zero below equation (\ref{eq:matt}) in the main text). The $N$ matrix is given by
\begin{equation}
N_{J^iJ^j}= N_{J^xJ^y}\,\epsilon^{ij}\, , \qquad
N_{J^iP^j}= -B\chi_{J^xJ^x}\,\epsilon^{ij}\, , \qquad
N_{P^iP^j}= -BQ\,\epsilon^{ij} \, .
\end{equation}

Inserting the above expressions in (\ref{eq:fullmemory2}) gives the following d.c.\ conductivities 
\begin{equation}
\sigma=0\,, \qquad
\sigma^H=\frac{Q}{B}\,.
\end{equation}
The optical conductivities have a pair of cyclotron poles at $\omega_\star = \pm\Omega_H-i\Omega$ with
\begin{subequations}\label{our_poles}
\begin{align}
\begin{split}
\ \ \Omega 
	&= \frac{B^2 \left(\chi_{J^xJ^x} - Q^2/\mathcal M\right)^2M_{J^xJ^x}}{\mathcal M (M_{J^xJ^x}^2 + (M_{J^xJ^y}+N_{J^xJ^y})^2)} \\
	&=\frac{B^2}{\mathcal M} \sigma_0+ \ocal(B^3)\, ,\\
\end{split}
\\
\begin{split}
\Omega_H
	&= \frac{Q B}{M}\left[1- \frac{B}{Q}\frac{ \left(\chi_{J^xJ^x} - Q^2/\mathcal M\right)^2(M_{J^xJ^y}+N_{J^xJ^y})}{\mathcal M (M_{J^xJ^x}^2 + (M_{J^xJ^y}+N_{J^xJ^y})^2)}
	\right]\\
	&=\frac{Q B}{\mathcal M}\left[1-\frac{B}{Q}\sigma_0^H\right]+ \ocal(B^3)\, ,\\
\end{split}
\end{align}
\end{subequations}
where we defined the incoherent conductivity matrix $(\sigma_0)^{ij} = \sigma_0 \,\delta^{ij}+ \sigma_0^H \epsilon^{ij}$ as the inverse of the incoherent resistivity matrix
\begin{equation}
(\rho_0)_{ij} = \frac{M_{J_i J_j} + N_{J_i J_j}}{\chi_{\rm inc}^2}\, ,
\end{equation}
with
\begin{equation}
\chi_{\rm inc} \equiv \chi_{J^xJ^x} - \frac{ Q^2}{\mathcal M}\, .
\end{equation}
These definitions are entirely analogous to those in section \ref{sec:memD}.

A clean result for the conductivities can be obtained with a scaling limit similar to (\ref{eq:scallim}), which is now expressed
\begin{equation}
\omega \sim B \, , \qquad
\chi_{\rm inc} \sim \frac{1}{\sqrt{B}}\, .
\end{equation}
The full conductivities in this limit are given by (typically $\sigma_0^H\sim B$, and hence it drops out of the final expression)
\begin{subequations}
\begin{align}
\sigma_{xx}(\omega)
	&=\frac{-i\mathcal M \omega (Q^2 + B^2 \sigma_0^2 - i\mathcal M\omega \sigma_0 )}{(-i\mathcal M \omega + \sigma_0 B^2)^2 + B^2 Q^2} + \ocal(B^0) \, ,\\
\sigma_{xy}(\omega)
	&=BQ\frac{ (Q^2 + B^2 \sigma_0^2 - 2i\mathcal M\omega \sigma_0 )}{(-i\mathcal M \omega + \sigma_0 B^2)^2 + B^2 Q^2} + \ocal(B^0) \, ,
\end{align}
\end{subequations}
which agrees with \cite{Hartnoll:2007ih, Hartnoll:2007ip, Lucas:2015pxa}.

\end{document}